\documentclass[a4paper,11pt]{article}
\usepackage{jheppub} 
\usepackage[T1]{fontenc} 
\usepackage{array}
\newcolumntype{P}[1]{>{\centering\arraybackslash}m{#1}}

\title{\boldmath Measurement of inclusive forward neutron production cross section in proton-proton collisions at $\sqrt{s} = 13$~TeV with the LHCf Arm2 detector}

\collaborationImg{\includegraphics[width=3.5cm]{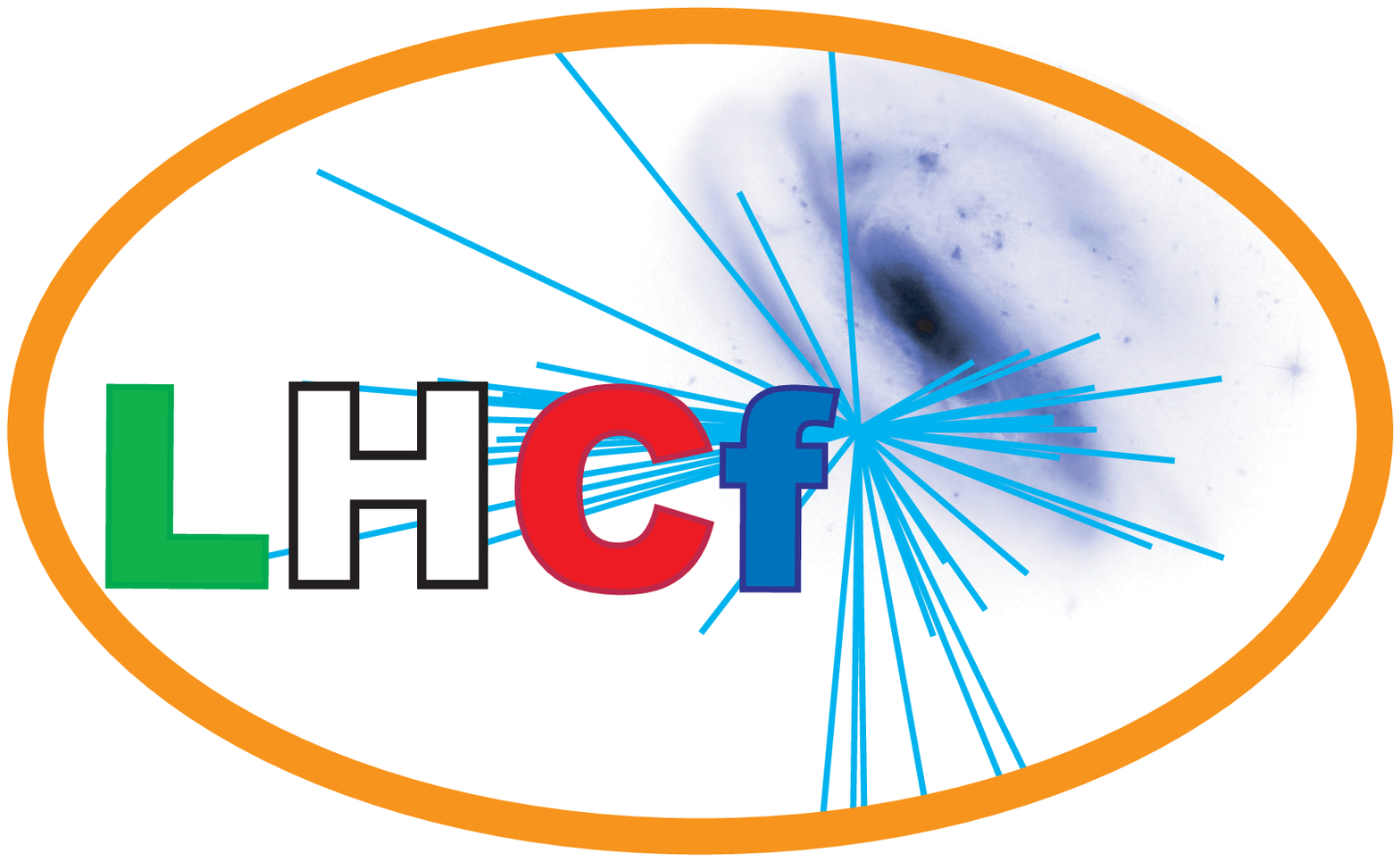} \newline \Large\bfseries\sffamily\raggedright{The LHCf collaboration} \afterCollaborationSpace}

\author[a,b]{O.~Adriani,}
\author[a,b,1]{E.~Berti,\note{Corresponding author.}}
\emailAdd{eugenio.berti@fi.infn.it}
\author[a]{L.~Bonechi,}
\author[a,b]{M.~Bongi,}
\author[a,b]{R.~D'Alessandro,}
\author[a]{S.~Detti,}
\author[c]{M.~Haguenauer,}
\author[d,e]{Y.~Itow,}
\author[f]{K.~Kasahara,}
\author[d,2]{Y.~Makino,\note{Present address: Department of Physics and Institute for Global Prominent Research, Chiba University, Chiba, Japan.}}
\author[d]{K.~Masuda,}
\author[g]{H.~Menjo,}
\author[d]{Y.~Muraki,}
\author[d]{K.~Ohashi,}
\author[a]{P.~Papini,}
\author[a,h]{S.~Ricciarini,}
\author[i]{T.~Sako,}
\author[j]{N.~Sakurai,}
\author[d]{K.~Sato,}
\author[d]{M.~Shinoda,}
\author[f]{T.~Suzuki,}
\author[k]{T.~Tamura,}
\author[a,b]{A.~Tiberio,}
\author[f]{S.~Torii,}
\author[l,m]{A.~Tricomi,}
\author[n]{W.C.~Turner,}
\author[d]{M.~Ueno,}
\author[d,3]{and Q.D.~Zhou\note{Present address: Institute of Particle and Nuclear Studies, High Energy Accelerator Research Organization (KEK), Tsukuba, Japan.}}

\affiliation[a]{INFN Section of Florence, Florence, Italy}
\affiliation[b]{University of Florence, Florence, Italy}
\affiliation[c]{Ecole-Polytechnique, Palaiseau, France}
\affiliation[d]{Institute for Space-Earth Environmental Research, Nagoya~University, Nagoya, Japan}
\affiliation[e]{Kobayashi-Maskawa Institute for the Origin of Particles and the Universe, Nagoya~University, Nagoya, Japan}
\affiliation[f]{RISE, Waseda University, Shinjuku, Tokyo, Japan}
\affiliation[g]{Graduate School of Science, Nagoya~University, Nagoya, Japan}
\affiliation[h]{IFAC-CNR, Florence, Italy}
\affiliation[i]{Institute for Cosmic Ray Research, University of Tokyo, Chiba, Japan}
\affiliation[j]{Tokushima University, Tokushima, Japan}
\affiliation[k]{Kanagawa University, Kanagawa, Japan}
\affiliation[l]{INFN Section of Catania, Italy}
\affiliation[m]{University of Catania, Catania, Italy}
\affiliation[n]{LBNL, Berkeley, California, USA}

\abstract{
In this paper, we report the measurement relative to the production of forward neutrons in proton-proton collisions at $\sqrt{s} = 13$~TeV obtained using the LHCf Arm2 detector at the Large Hadron Collider. The results for the inclusive differential production cross section are presented as a function of energy in three different pseudorapidity regions: $\eta > 10.76$, $8.99 < \eta < 9.22$ and $8.81 < \eta < 8.99$. The analysis was performed using a data set acquired in June 2015 that corresponds to an integrated luminosity of $\mathrm{0.194~nb^{-1}}$. The measurements were compared with the predictions of several hadronic interaction models used to simulate air showers generated by Ultra High Energy Cosmic Rays. None of these generators showed good agreement with the data for all pseudorapidity intervals. For $\eta > 10.76$, no model is able to reproduce the observed peak structure at around 5~TeV and all models underestimate the total production cross section: among them, QGSJET II-04 shows the smallest deficit with respect to data for the whole energy range. For $8.99 < \eta < 9.22$ and $8.81 < \eta < 8.99$, the models having the best overall agreement with data are SIBYLL 2.3 and EPOS-LHC, respectively: in particular, in both regions SIBYLL 2.3 is able to reproduce the observed peak structure at around 1.5-2.5~TeV.
}

\keywords{LHCf, Large Hadron Collider, Hadronic Interaction Models, Ultra High Energy Cosmic Rays, Forward Leading Baryon}

\begin{document} 
\maketitle
\flushbottom

\section{Introduction}

In recent years, several ground-based experiments, like the Pierre Auger Observatory \cite{ref:PAO} and Telescope Array \cite{ref:TA}, have measured the flux and composition \cite{ref:PAO_Xmax, ref:TA_Xmax} of cosmic rays at energies above $10^{18}$~eV, the Ultra High Energy Cosmic Rays (UHECRs). In all cases, the results are affected by large systematic uncertainties because of the reliance on hadronic interaction models necessary for the data analysis. These generators are in fact extensively used in the reconstruction of incident cosmic rays starting from the Extensive Air Shower (EAS) they form when interacting in the atmosphere. Because of the lack of calibration data, relative to forward secondary particle production in high energy collisions, these models show a very different behavior, especially in the phase space regions that are relevant to the EAS physics \cite{ref:Ulrich}. Some of these generators, namely QGSJET II-04 \cite{ref:QGSJET}, EPOS-LHC \cite{ref:EPOS} and SIBYLL 2.3 \cite{ref:SIBYLL}, were tuned using the measurements made in p-p collisions at $\sqrt{s} = 0.9$ and 7~TeV, during the Run I operations at CERN Large Hadron Collider (LHC) \cite{ref:LHC}, the highest energy accelerator currently available. Even considering these post-LHC models, the level of agreement between them is far from satisfactory \cite{ref:PAO_model} and higher energy data is needed to reduce the observed discrepancies. Among LHC experiments, the LHCf experiment \cite{ref:LHCf_TDR} has been designed to specifically measure the distributions of neutral particles produced in the very forward region in p-p and p-nucleus collisions. Of particular interest are the forward neutral hadrons detectable by the experiment, mainly neutrons, because of the key role that forward baryons have in the development of the atmospheric showers and in the abundance of the muonic component \cite{ref:Pierog}. The measurements relative to the inclusive differential cross section of forward neutrons produced in p-p collisions at $\sqrt{s} = 7$~TeV have already been published by our collaboration \cite{ref:LHCf_7TeV}. In this paper, we extend this analysis to the data obtained in p-p collisions at $\sqrt{s} = 13$~TeV, making use of the same pseudorapidity regions defined in the previous work. Compared to the 7~TeV run, this corresponds to an effective increase of the energy in the laboratory frame by a factor of four and an increase in the $\mathrm{p_{T}}$ coverage by a factor of two. The results presented here were obtained using data from only one of the two LHCf detectors (Arm2), but in the future we will extend this analysis to the other one (Arm1) in order to enlarge the pseudorapidity coverage, exploiting the slightly different geometrical acceptance of the two detectors.

\section{The LHCf experiment}

The LHCf experiment consists of two detectors \cite{ref:LHCf_detector}, Arm1 and Arm2, placed in two regions on the opposite sides of LHC Interaction Point 1 (IP1). These regions, called Target Absorber Neutral (TAN), are located at a distance of 141.05~m from IP1, after the dipole magnets that bend the two proton beams. In this position, the two detectors are capable of detecting neutral particles produced in p-p and p-nucleus collisions with a pseudorapidity coverage $\eta > 8.4$. 

Arm2 consists of two calorimetric towers made from 16 $\mathrm{Gd_{2}SiO_{5}}$ (GSO) scintillators layers (1~mm thick) interleaved with 22 tungsten (W) plates (7~mm thick) for a total length of about 21~cm, equivalent to 44~$\mathrm{X_{0}}$ and 1.6~$\mathrm{\lambda _{I}}$. The transverse sizes of the two towers, called \textit{small tower} and \textit{large tower}, are respectively 25~mm $\times$ 25~mm and 32~mm $\times$ 32~mm. Apart from the energy, measured using the 16 scintillator layers, the transverse position is also reconstructed from the transverse profile of the showers, measured using 4 xy imaging layers. These imaging layers, placed at different depths in the calorimeter, consist of $\mathrm{160~\mu m}$ read-out pitch silicon microstrip detectors. 

\enlargethispage{+2\baselineskip}
The detector just illustrated is an upgraded version of the one described in reference \cite{ref:LHCf_7TeV}: the upgrade was necessary to improve the radiation hardness of the detector for the LHC Run II operations. However, the performance of the current detector in the reconstruction of hadronic showers, investigated making use of beam test data and MC simulations, are not significantly different from the old ones \cite{ref:LHCf_performances}. For hadrons between 1 and 6.5~TeV, detection efficiency ranges from about 52\% to 72\%, energy resolution from 28\% to 38\%, and position resolution from $\mathrm{300~\mu m}$ to $\mathrm{100~\mu m}$.

\section{Experimental data set}
\label{sec:exp}

The data analyzed in this paper was acquired during p-p collisions at $\sqrt{s} = 13$~TeV, from 22:32 of June 12th to 1:30 of June 13th 2015 (CEST). This time period corresponds to the LHC Fill 3855, a special low luminosity and high $\beta^{*}$ fill specifically provided for LHCf operations. In this fill, 29 bunches collided at IP1 with a half crossing angle of 145~$\mathrm{\mu rad}$ downward. Additionally, 6 and 2 bunches that did not collide at IP1 circulated in the clockwise and counter-clockwise beams, respectively. Beam $\beta^{*}$ was around 19~m and $\mu$, the average number of collisions per bunch crossing, was in the range 0.007--0.012. Given the 15\% acceptance of the calorimeter for inelastic collisions, event pile-up at the detector level was estimated to be below 1\%, and was therefore neglected in this analysis. The number of shower events in Arm2 that were recorded during this time period was $2.1\times10^{6}$. Using the instantaneous luminosity measured from the ATLAS experiment (3-$\mathrm{5\times 10^{28}~cm^{-2} s^{-1}}$, \cite{ref:ATLAS-luminosity}) and taking into account LHCf data acquisition live time, the recorded integrated luminosity was estimated to be $\mathrm{0.194~nb^{-1}}$. 

\section{Simulation data set}
\label{sec:mc}

Monte Carlo simulations (MC) with the same experimental configuration used for LHC data taking are necessary for four different purposes: estimation of correction factors and systematic uncertainties, validation of the whole analysis procedure, energy spectra unfolding and comparison of the models with the experimental results. All simulations steps (collision, transport and interaction with the detector) used Cosmos 7.633 \cite{ref:COSMOS} and EPICS 9.15 \cite{ref:EPICS} libraries, except where otherwise specified. In the following, MC are separated in three different categories, summarized in table \ref{tab:simulations}: 

\begin{itemize}
\enlargethispage{+2\baselineskip}
\item[\textit{a})] MC that require a full detector simulation, i.e. generating collisions, transporting secondary products from IP1 to the TAN region (taking into account the effect of the magnetic field, particle decays and particle interactions with the beam pipe) and finally simulating the interaction with the detector. Three different models were used to generate collisions: QGSJET II-04, EPOS-LHC and DPMJET 3.04 \cite{ref:DPMJET}, with a corresponding statistics of $10^{8}$,  5$\times 10^{7}$ and 2$\times 10^{7}$ inelastic collisions, respectively. In all cases, the transport along the beam pipe and the interaction with the detector were simulated using the DPMJET 3.04 model. 
In addition, we generated another special set of 4$\times 10^{7}$ inelastic collisions dedicated to multihit events (events in which more than one particle enters a given calorimetric tower) both for QGSJET II-04 and EPOS-LHC. These special events were generated in two steps. First, starting from the full sample described above, we separated each multihit event, where \textit{n} particles simultaneously hit the detector, into \textit{n} independent singlehit events (events in which only one particle enters a given calorimetric tower). Then, we simulated again the interaction with the detector, this time one particle at a time.  All the full detector simulations were used for the validation of the whole analysis procedure and for a preliminary comparison with experimental data at the folded spectra level, as well as for the estimation of a part of the correction factors (beam-pipe background, Particle IDentification (PID), multihit, fake, missed, see section \ref{sec:corrections}) and of the systematic uncertainties (position resolution, unfolding method, see section \ref{sec:systematics}). Note that for this last purpose (evaluation of corrections and uncertainties), only QGSJET II-04 and EPOS-LHC were used as collision generators, whereas DPMJET 3.04 was excluded because of its very limited agreement with data already at the folded spectra level (as discussed in section \ref{sec:results}).

\begin{table} [tbp]
  \scriptsize
  \begin{center}
    \begin{tabular}{|P{1.5cm}|P{1.5cm}|P{2.0cm}|P{1.5cm}|P{2.0cm}|P{1.5cm}|P{2.0cm}|}
     \hline
      & \multicolumn{2}{c}{} & \multicolumn{2}{|c|}{} & \multicolumn{2}{c|}{} \\
      \textbf{Simulation} & \multicolumn{2}{c}{\textbf{Collisions generation}} & \multicolumn{2}{|c|}{\textbf{Particles Transport}} & \multicolumn{2}{c|}{\textbf{Detector interaction}} \\
      & \multicolumn{2}{c}{} & \multicolumn{2}{|c|}{} & \multicolumn{2}{c|}{} \\
      \hline
      \vspace{1cm} & \vspace{1cm} & \vspace{1cm} & \vspace{1cm} & \vspace{1cm} & \vspace{1cm} & \vspace{1cm}\\
      & \textbf{Interface}  & \textbf{Model} & \textbf{Interface} & \textbf{Model} & \textbf{Interface} & \textbf{Model}\\
      \vspace{1cm} & \vspace{1cm} & \vspace{1cm} & \vspace{1cm} & \vspace{1cm} & \vspace{1cm} & \vspace{1cm}\\
      \hline
	  \textit{a}	& Cosmos-EPICS  & QGSJET II-04, EPOS-LHC, DPMJET 3.04	& Cosmos-EPICS &  DPMJET 3.04 & Cosmos-EPICS & DPMJET 3.04\\
      & & & & & & \\      
	  \textit{b}	& CRMC & QGSJET II-04, EPOS-LHC, SIBYLL 2.3, DPMJET 3.06, PYTHIA 8.212	& Cosmos-EPICS & DPMJET 3.04 & - & -\\
      & & & & & & \\	  
	  \textit{c}	& -	 & - & -  & - & Cosmos-EPICS & DPMJET 3.04, QGSJET II-04\\
      \hline
    \end{tabular}
    \caption{Summary of the three MC sets described in the text. The simulation is divided in three main steps: generation of particles produced in p-p collisions at $\sqrt{s} = 13$~TeV, transport of secondary products from IP1 to the TAN region (taking into account the effect of the magnetic field, particle decays and particle interactions with the beam pipe), interaction of particles with the Arm2 detector. For each MC set we specified which software interfaces and which interaction models were used to simulate each step. The symbol ``-'' indicates that it was not necessary to simulate the given step for that MC set.
    }
    \label{tab:simulations}
  \end{center}
\end{table}

\item[\textit{b})] MC that require only the generation of collisions and the transport of secondary products from IP1 to the TAN region, without taking into account the interaction with the detector. We simulated $10^{8}$ collisions for each of the following models: QGSJET II-04, EPOS-LHC, SIBYLL 2.3, DPMJET 3.06 and PYTHIA 8.212  \cite{ref:PYTHIA}. The PYTHIA 8.212 simulation data set was generated using its own dedicated generator, whereas CRMC \cite{ref:CRMC}, an interface tool for event generators, was employed for the remaining models (version 1.5.6 for DPMJET 3.06, version 1.6.0 for the others). In both cases, all particles having $\mathrm{c \tau < 1~cm}$ were considered unstable and their decay was treated by the interface. After generation, the secondary products were transported along the beam pipe making use of the DPMJET 3.04 model. All these simulations were used in the estimation of the hadron contamination correction factors and for the comparison with the measured distributions after unfolding.

\enlargethispage{+1\baselineskip}
\item[\textit{c})] MC that require only the interaction with the detector to construct the response matrix used for unfolding. In order to be independent of the particular shape of the injected distribution, we considered a single neutron flat energy spectrum from 0 to 6.5~TeV, uniformly distributed on the whole area of the detector. The interaction with the detector was simulated using DPMJET 3.04 and QGSJET II-04, so that we could estimate the dependence of the results on the model adopted in the unfolding procedure.\footnote{
In the case of QGSJET II-04, the simulation of hadronic showers in the calorimeter was carried out using DPMJET 3.04 for particles below 90 GeV.
} This sample consists of about $10^{7}$ neutrons per tower for each model.
\end{itemize}
As a curious aside, we found that in case of p-p collisions at $\sqrt{s} = 13$~TeV, QGSJET II-04 predicts a significant number of neutrons having exactly $\mathrm{p_{T}}$ = 0~GeV/c. According to the author's explanation \cite{ref:Ostapchenko}, these neutrons arise when a valence quark from one of the two protons is involved in a collision having a small impact parameter. For simplicity, a temporary value of $\mathrm{p_{T}}$ = 0~GeV/c was assigned during model implementation, but in fact these events should not contribute to particle production in the forward region. Because of this reason, following the author's suggestion, in the current analysis such kind of neutrons was removed a posteriori from all simulation sets relative to this generator.

\section{Analysis}

In this section, we present the analysis procedure, divided in four main steps. First, after defining the event selection criteria, we reconstruct the events in the sample and derive the folded energy distributions for each pseudorapidity region (section \ref{sec:reconstruction}). Second, we estimate the correction factors that are then applied to the folded spectra (section \ref{sec:corrections}). Third, we unfold the energy distributions using a deconvolution procedure that takes into account the detector response (section \ref{sec:unfolding}). Fourth, all necessary systematic uncertainties are evaluated and associated to the folded and unfolded spectra (section \ref{sec:systematics}). Finally, in the next section, we will use these distributions to derive the inclusive differential production cross section and compare it to the various model predictions (section \ref{sec:results}).

\subsection{Event reconstruction}
\label{sec:reconstruction}

The event reconstruction algorithm used in this analysis is similar to the one developed in reference \cite{ref:LHCf_7TeV}. Because of the detector upgrade after the LHC Run I operations, we revised all the calibration parameters and optimized the reconstruction criteria, making use of beam test and MC simulation data. In particular, because of the inherent difficulty found in correctly identifying a multihit hadronic event, we decided to reconstruct all events as singlehit and to apply afterwards a correction factor, as described in section \ref{sec:corrections}.\footnote{
Reconstructing an event as singlehit means that we always identify only one single peak in the transverse shower profile as imaged by the silicon layers, independently on the number of peaks actually present.
}

\enlargethispage{-2\baselineskip}

An event is accepted if it satisfies a software trigger condition, defined as a raw energy deposit above 850~MeV in at least three consecutive scintillator layers. This value was chosen by taking into account the efficiency curves of the discriminators used to generate the hardware trigger during LHC operations. Since hadron detection efficiency falls sharply for true hadron incident energies below 500~GeV, we report our final result only above this value. However, in all analysis steps we considered all reconstructed energies over 250~GeV, because we found that this choice improves the unfolding accuracy on the final result. The reason is that, due to the limited hadron energy resolution of the detector, low energy tails in the deposits of high energy hadrons lead to a significant contribution in the 250--500~GeV range of the folded distribution.

\begin{figure*}[tbp]
 \centering
 \includegraphics[width=0.5\textwidth]{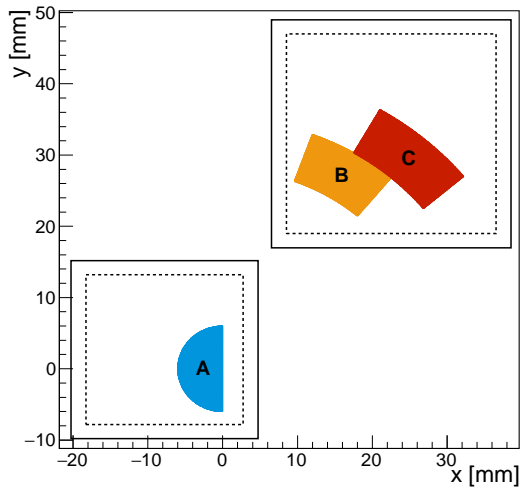}
 \caption{ 
  Definition of the three pseudorapidity regions A (blue), B (yellow) and C (red) on the Arm2 detector seen from IP1. The origin of the reference frame is centered on the projection of beam center on the detector during Fill 3855. The left and the right squares correspond to the small tower and the large tower, respectively. All analysis regions are chosen within a fiducial area (dashed line), which is 2~mm inside the edges of the towers (solid line).   
  }
 \label{fig:regions}
\end{figure*}

In this analysis, we used the same three pseudorapidity regions defined in reference \cite{ref:LHCf_7TeV}, here called A, B and C, that correspond to a $\eta$ coverage of $\eta > 10.76$, $8.99 < \eta < 9.22$ and $8.81 < \eta < 8.99$, respectively. Figure \ref{fig:regions} shows these regions superimposed on the detector area (more details are given in table \ref{tab:regions}). Due to the limited position resolution of the imaging layers, a small migration of events inside and outside the pseudorapidity regions is present. This effect is taken into account both by including it in the fake and missed events correction factors and by adding a position resolution systematic uncertainty. 

\begin{table}[tbp]
  \begin{center}
    \begin{tabular}{|c|c|c|c|}
      \hline
      & \textbf{Region A} & \textbf{Region B} & \textbf{Region C} \\
      \hline
	  $\eta$	& 10.76--$\infty$	& 8.99--9.22	 & 8.81--8.99\\
	  $r~\mathrm{[mm]}$	& 0--6	& 28--35	& 35--42\\
	  $\Delta \phi~[\ensuremath{^\circ}]$	& 180	& 20 & 20\\
      \hline
    \end{tabular}
    \caption{Definition of the three pseudorapidity regions used in the analysis: $\eta$ is the pseudorapidity,  $r$ is the distance from the beam center in the detector plane and $\Delta \phi$ is the azimuthal angle coverage.}
    \label{tab:regions}
  \end{center}
\end{table}

\enlargethispage{+1\baselineskip}

Particle identification exploits the variable $L_{2D} = L_{90\%} - 0.25 \times L_{20\%}$, $L_{20\%}$ and $L_{90\%}$ representing the longitudinal depths where the fraction of the energy deposit of a shower  compared to the total release in the calorimeter is $20\%$ and $90\%$, respectively. For each pseudorapidity region, we identified as neutrons all events having $L_{2D} > L_{2D}^{thr}$, where $L_{2D}^{thr}$ is a threshold value, estimated from simulations, that maximizes the product of efficiency vs purity. Although this threshold has in principle a slight energy dependence, for simplicity we chose a constant value for all energy bins within the same pseudorapidity region, and later applied a correction for residual inefficiencies and contaminations.

\subsection{Correction factors}
\label{sec:corrections}

\begin{figure*}[tbp]
 \centering
 \includegraphics[width=.333\textwidth]{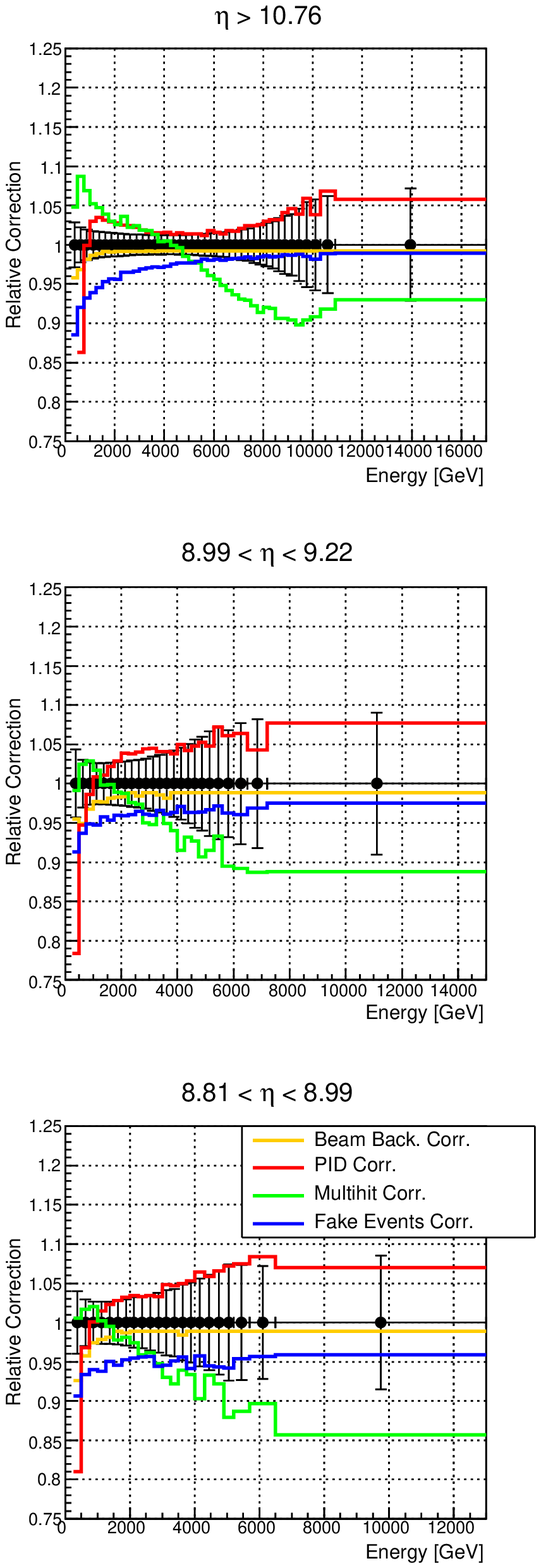}%
 \includegraphics[width=.333\textwidth]{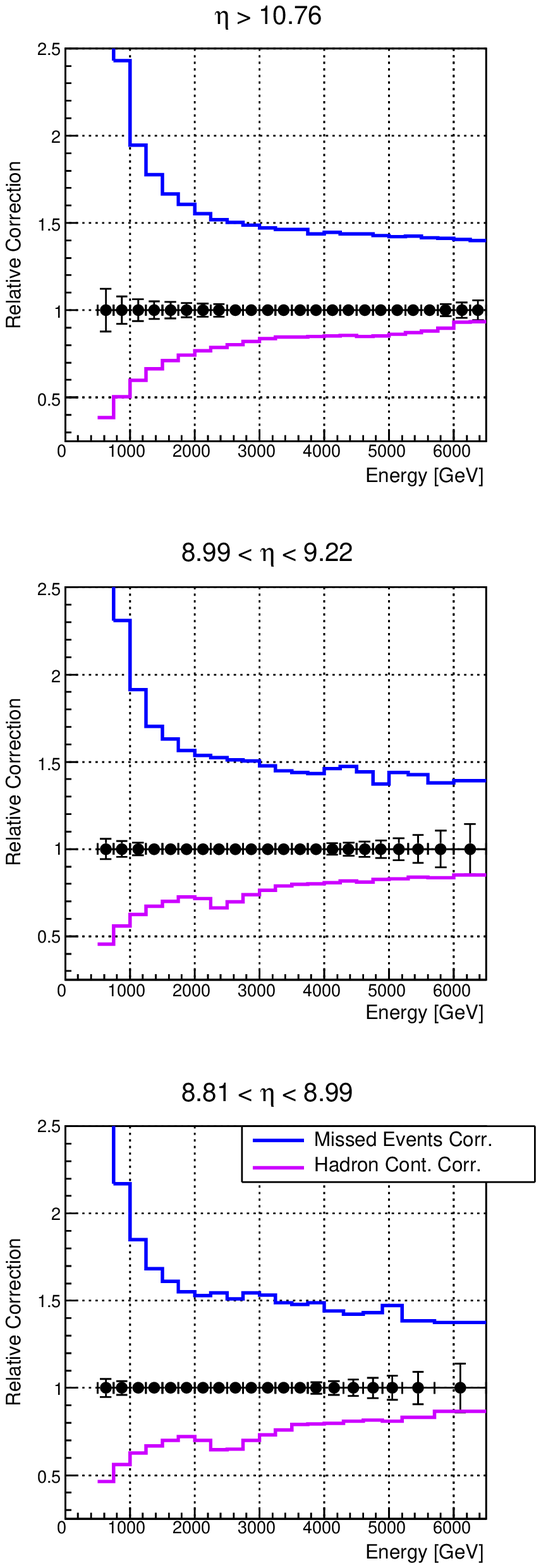}%
 \caption{Relative correction factors before (left) and after (right) unfolding. In the two cases, black markers show the statistical error present on the folded and the unfolded distributions, respectively. The different colors represent the relative change of the histograms after each correction.}
\label{fig:corrections}
\end{figure*} 

\begin{table}[tbp]
  \begin{center}
    \begin{tabular}{|c|c|c|c|}
       	\hline
		\textbf{Correction} & \textbf{Region A} & \textbf{Region B} & \textbf{Region C} \\
    	\hline
		Beam background 	  &	 (-5\%, -1\%)  	&	 (-5\%, -1\%)   & (-5\%, -1\%) \\
		PID 	  &	 (+1\%, +6\%) 	&	 (-5\%, +8\%)   & (-5\%, +8\%) \\
		Multihit 	  &	 (-10\%, +9\%) 	&	 (-12\%, +3\%)   & (-14\%, +2\%) \\
		Fake events   &  (-8\%, -1\%) 	&	 (-7\%, -2\%)  & (-7\%, -4\%) \\
		Missed events &  (+40\%, +270\%)   &  (+40\%, +240\%)  & (+40\%, +230\%) \\      
		Hadron contamination &  (-60\%, -7\%)   &  (-55\%, -15\%)  & (-55\%, -15\%) \\      
		\hline
    \end{tabular}
	\caption{(Minimum, maximum) value of the relative correction factors for each contribution discussed in section \ref{sec:corrections}, separately reported for pseudorapidity regions A, B and C.}
    \label{tab:corrections}
  \end{center}
\end{table}

\enlargethispage{+1\baselineskip}
The values of correction factors estimated for the current analysis are summarized in figure \ref{fig:corrections} and table \ref{tab:corrections}. Apart from beam-gas contamination, all of them were calculated independently for each energy bin. Corrections were applied to the energy spectra, before or after unfolding, in the same order they are mentioned in the following:

\begin{itemize}
\item Beam background

Beam-related background is due to two different processes: the interaction of primary protons with residual gas in the beam line and the interaction of secondary particles from collisions with the beam pipe. The first contribution was estimated using events associated with non-crossing bunches at IP1, generated purely from beam-gas interactions, and comparing this number to the events associated with colliding bunches, that include both signal and background. We obtained a background-to-signal ratio of about 1\%. The second contribution was estimated making use of MC simulations in which we simulated the interaction with the beam pipe, leading to a background to signal ratio that depends on the reconstructed energy, ranging from about 5\% at 500~GeV to less than 1\% above 1~TeV. Both corrections were applied to the measured distributions and the uncertainty was neglected, being always smaller than 1\%.

\item PID

PID has a limited efficiency (hadron identification) and purity (photon contamination). Correction factors were estimated by performing a template fit of the $L_{2D}$ distributions of photons and hadrons, derived from simulations, to the $L_{2D}$ distributions experimentally observed. An example of this procedure is shown in figure \ref{fig:PID}. We did not observe any significant dependence on the generator, since the correction factors are more related to the detector response than to the distributions of incoming particles. Thus, the uncertainty on the PID correction factors is mainly statistical, estimated using a confidence interval method on the result of the template fit. The correction ranges between -5\% and +8\%, whereas the uncertainty is below 10\%.

\begin{figure*}[tbp]
 \centering
 \includegraphics[width=0.5\textwidth]{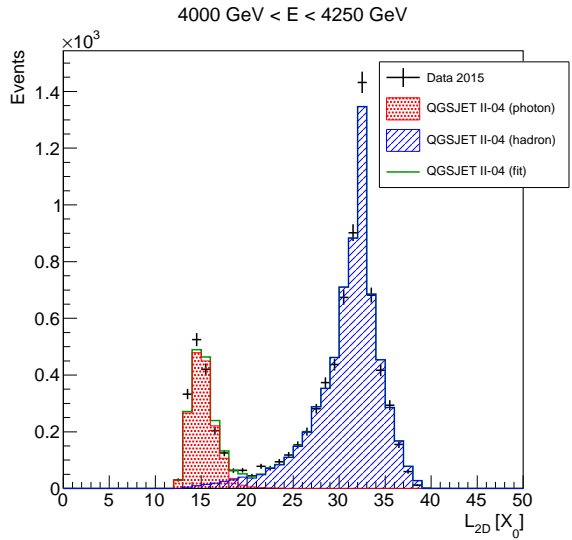}
 \caption{Template fit for the reconstructed energy bin $\mathrm{4000~GeV < E < 4250~GeV}$ of pseudorapidity region A. The binning of the $L_{2D}$ scale varies depending on the expected statistics for each energy bin. QGSJET II-04 hadrons (blue) and photons (red) distributions were fitted to experimental data (black). The result of the fit is shown in green.}
\label{fig:PID}
\end{figure*}

\item Multihit

Multihit events are not correctly identified by the reconstruction algorithm described in section \ref{sec:reconstruction}. For this reason, we estimated the relative correction factors using the special simulation sample described in section \ref{sec:mc}: in this MC, each multihit event involving $n$ particles simultaneously hitting the same tower was subdivided in $n$ different events, one per particle. This sample was reconstructed in two different ways: in one case we artificially piled-up each group of $n$ events in a single multihit event, whereas in the other case we considered each event individually. These two cases correspond, respectively, to the real/ideal situation in which our detector cannot/can distinguish multihit events. Corrections were calculated making use of the QGSJET II-04 and the EPOS-LHC generators, with the final correction factors estimated from the average over the two models and the systematic error given by half their difference. The correction ranges between -14\% and +9\%, whereas the uncertainty is below 10\%.

\item Fake and missed

\textit{Fake} and \textit{missed} events refer, respectively, to events that, due to detector misreconstructions, are either incorrectly \textit{included in}, or incorrectly \textit{excluded from}, the measured distributions. Both effects are due to improper position and/or energy reconstruction, which are significant only in the low energy region. As discussed in section \ref{sec:reconstruction}, the limited hadron detection efficiency contributes as well to the missed events. Actually, it constitutes the largest contribution to correction factors: indeed, due to the small depth of the detector, a large fraction of hadrons either interacts late or does not interact at all in the calorimeter. Above 2 TeV, the detection efficiency is mostly constant around a 70\% value, whereas, below 2 TeV, it strongly decreases due to the smaller energy deposits left in the scintillators. Note that fake and missed events corrections are expressed as a function of different definitions of energy (reconstructed and true) because they are respectively applied before and after unfolding. They were both estimated making use of the ideal (multihit corrected) sample described before, using the average between QGSJET II-04 and EPOS-LHC as the final correction factors. Since different generators change the result by less than 1\%, we decided to neglect the contribution to the total systematic uncertainty due to model dependence of these correction factors. Fake events correction (i.e. $1-f_{i}$, where $f_{i}$ is the ratio between the entries that \textit{should not be} reconstructed in bin $i$ and the number of reconstructed events in bin $i$) slightly decreases with increasing energy, ranging from -8\% to -1\%. Missed events correction (i.e. $m_{j}$, where $m_{j}$ is the ratio between the entries that \textit{should be} reconstructed in bin $j$ and the number of reconstructed events in bin $j$) ranges from more than +100\% at 500~GeV to about +40\% at 6.5~TeV. 

\item Hadron contamination

A non-negligible fraction of hadrons reaching the detector are not actually neutrons, but mainly $\Lambda^{0}$, $\mathrm{K^{0}_{L}}$ and other neutral and charged hadrons. Since the detector does not distinguish between them, the unfolded spectra must be corrected in order to recover the distributions of neutrons at IP1. \textit{Neutrons at IP1} include both neutrons produced from p-p collisions and neutrons produced from the decay of short lived particles ($\mathrm{c \tau < 1~cm}$). The hadron contamination was calculated using all five generators described in section \ref{sec:mc}, also simulating the transport of the secondary products from IP1 to the TAN region. The correction factors were estimated from the average over the five models with a systematic error given by the maximum deviation observed. The correction ranges between -60\% and -7\%, whereas the uncertainty is below 20\%.

\end{itemize}

\subsection{Spectra unfolding}
\label{sec:unfolding}

Given the limited hadron energy resolution, we chose to perform an unfolding of the spectra in order to deconvolute our measurements from the detector response. For this purpose, we used the iterative Bayesian method \cite{ref:DAgostini}, implemented in the RooUnfold package \cite{ref:RooUnfold}. Some changes to the default implementation were made, in order to separate the choice of the prior used as initial hypothesis of the algorithm from the distribution employed to construct the response matrix of the detector. This modification was necessary because we need to distinguish between the different contributions to the systematic uncertainty coming from the unfolding procedure itself, as discussed in section \ref{sec:systematics}. The best estimation for the final unfolded distributions was obtained choosing a flat prior as initial hypothesis and constructing the response matrix from the single neutron flat energy sample simulated using the DPMJET 3.04 model. In order to estimate the uncertainties related to the deconvolution, the unfolding procedure was repeated several times for different priors, response matrices, or test input spectra. It was therefore important to define a general convergence criteria that could be valid for all the different cases tested: we decided to stop the iterative procedure when $\Delta \chi^{2}$, the $\chi^{2}$ change between the outputs of two consecutive iterations, is below 1. This threshold was chosen as a compromise between convergence requirements and number of iterations needed to reach that convergence. As an example, in the case of the data points in the final unfolded distributions, this convergence criteria corresponds to 51, 12 and 10 iterations for Region A, B and C, respectively.
Differently from the default procedure, we also chose to apply the fake and missed events corrections outside of the unfolding algorithm, although we verified that the results are consistent with the standard approach. Note that this procedure is different from the one used in reference \cite{ref:LHCf_7TeV}.

\subsection{Systematic uncertainties}
\label{sec:systematics}

\begin{figure*}[tbp]
 \centering
 \includegraphics[width=.333\textwidth]{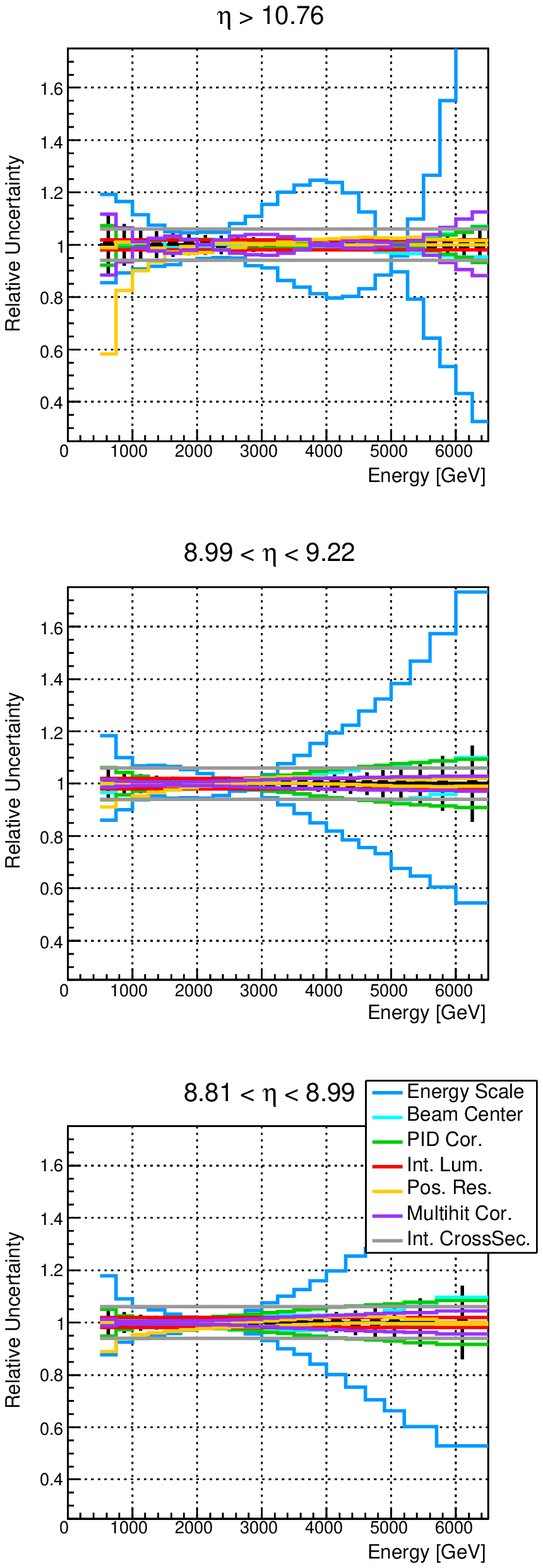}%
 \includegraphics[width=.333\textwidth]{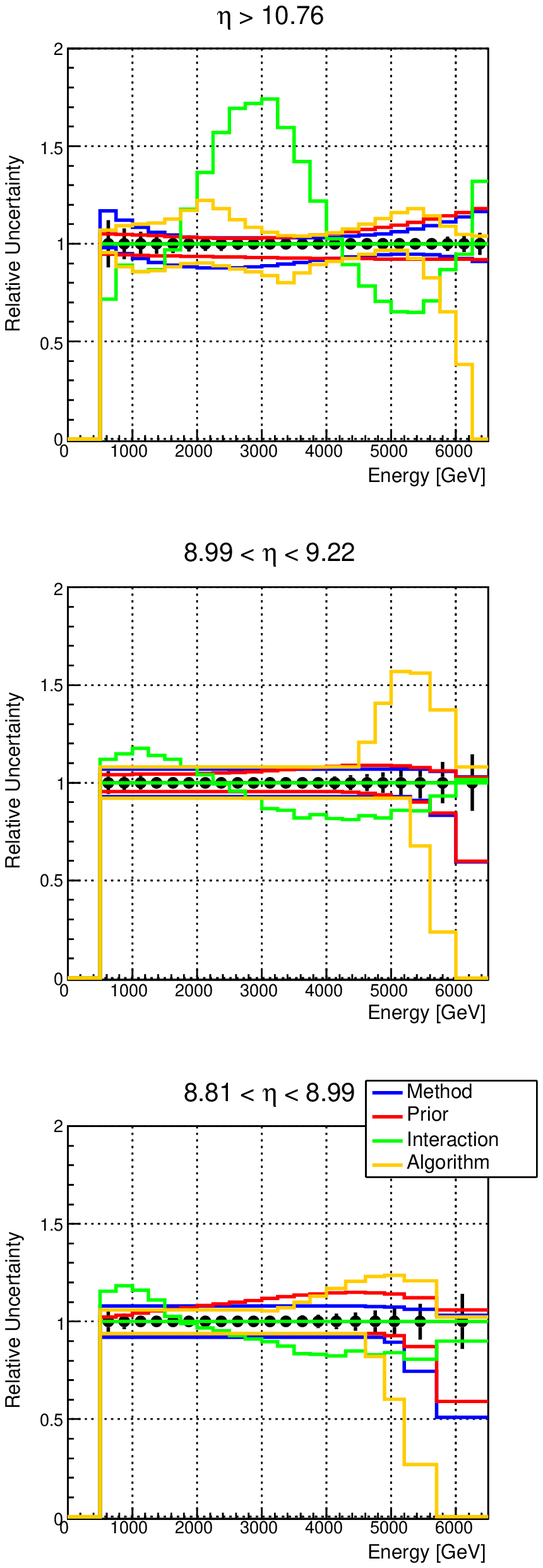}%
 \includegraphics[width=.333\textwidth]{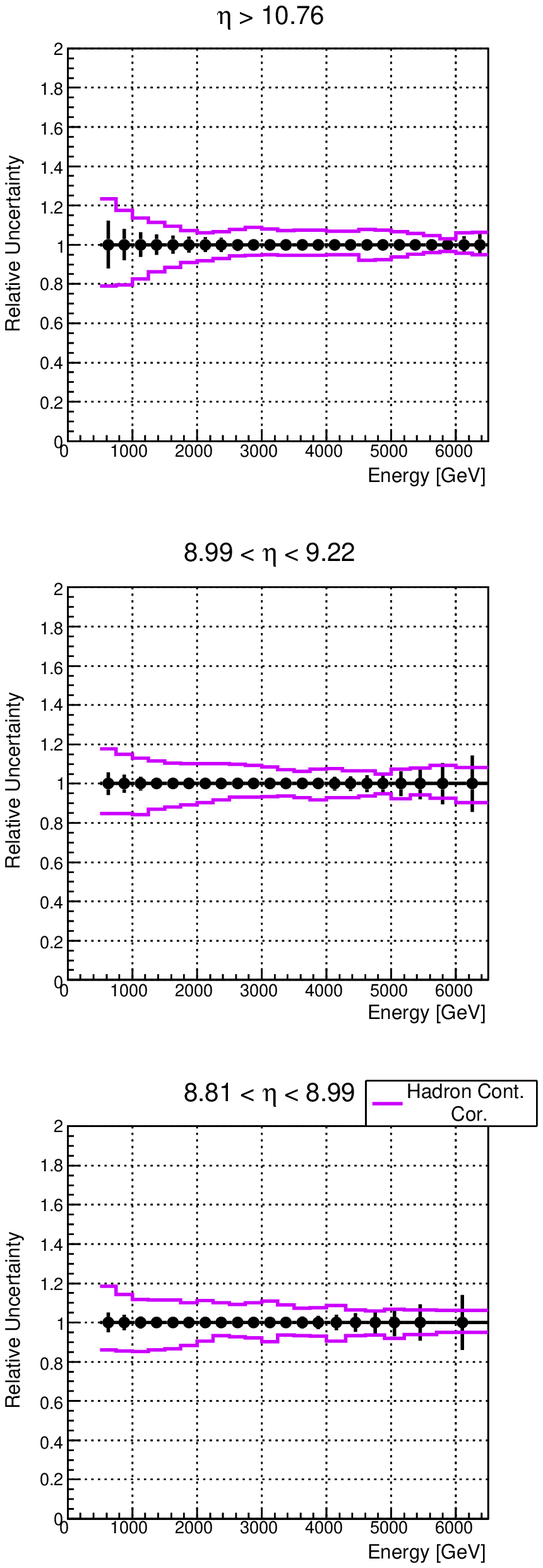}%
 \caption{
 Relative systematic uncertainties for the final unfolded distributions. From left to right, the three columns represent: the uncertainties already associated to the folded spectra, the uncertainties due to the unfolding procedure and the uncertainties related to hadron contamination corrections. Black markers show the statistical error present on the unfolded distribution.}
\label{fig:systematics}
\end{figure*}

\begin{table}[tbp]
  \begin{center}
    \begin{tabular}{|c|c|c|c|}
       	\hline
		\textbf{Systematic} & \textbf{Region A} & \textbf{Region B} & \textbf{Region C} \\
    	\hline
		Energy scale 	& 1--130\% &	 1--70\% & 	 1--60\%  \\
		Beam center 	& 1--7\% &	 1--10\% &	 1--10\%  \\
		Position resolution 	 	& 1--40\% & 	 1--10\% &	 1--10\%  \\
		Integrated luminosity 	 	& 2\% &	 2\% &	 2\%  \\
		PID correction 	 	& 1--8\% &	 1--9\% &	 1--9\%  \\
		Multihit correction 	 	& 1--10\% &	 1--3\% &	 1--3\%  \\
		Interaction cross section 	 	& 6\% &	 6\% &	 6\%  \\
		Unfolding method 	 	& 3--20\% &	 3--40\% &	 3--50\%  \\
		Unfolding prior 	 	& 3--20\% &	 3--40\% &	 3--40\%  \\
		Unfolding interaction 	 	& 1--70\% &	 1--20\% &	 1--20\%  \\		
		Unfolding algorithm 	 	& 3--100\% &	 8--100\% &	 3--100\%  \\
		Hadron contamination correction 	 	& 3--20\% &	 5--20\% &	 5--20\%  \\				      
		\hline
    \end{tabular}
	\caption{Minimum-maximum value of the relative systematic uncertainties for each contribution discussed in section \ref{sec:systematics}, separately reported for pseudorapidity regions A, B and C.}
    \label{tab:systematics}
  \end{center}
\end{table}

The typical value of systematic uncertainties estimated for the current analysis are summarized in figure \ref{fig:systematics} and table \ref{tab:systematics}. All of them were assumed to be independent and added in quadrature in the final result. Attention must be paid in distinguishing between uncertainties associated to the spectra after unfolding (unfolding error, hadron contamination) and before unfolding (all the remaining contributions). While the first can simply be added to the final distributions, the second must be correctly propagated through the unfolding procedure. Since the iterative Bayesian unfolding is capable of handling statistical but not systematic errors, we devised the following procedure. First, we considered each contribution independently and artificially shifted the folded nominal spectrum by its estimated systematic uncertainty. Then, we unfolded the two distributions, that corresponds to the two extremes of the systematic error bars, using the same number of iterations as for the nominal case. Lastly, we derived the systematic uncertainty on the unfolded distribution from the ratio between the unfolded spectra relative to the error bar and the unfolded nominal one. In the following we detail all contributions to the systematic uncertainties:

\begin{itemize}
\item Energy scale

The source of uncertainty on the energy scale comes from three different contributions. The first is due to calorimeter calibration (3.7\%), comprising the absolute gain calibration for each scintillator layer, the non-uniformity of the position dependent correction factors, and the residuals in the conversion from the deposited energy to primary energy. The second comes from different hardware effects present during data taking at the LHC (2.0\%), like ADCs non-linearity, cables attenuation, high voltages stability, PMTs gain calibration. The third is related to the observed shift in the position of the $\pi^{0}$ mass peak with respect to the simulations (+1.6\%), as reconstructed from the two gamma invariant mass distribution \cite{ref:LHCf_13TeV}. Despite the fact that this shift is well within the consistency bounds of the two previously mentioned systematic uncertainties, we decided to consider this contribution as an additional source of error in our analysis. This conservative approach is justified, in the case of hadronic showers, because we do not have a clear energy signature that can be used to calibrate the absolute energy scale. The total energy scale error, obtained from the quadratic sum of these three terms, is $\pm 4.5\%$. The systematic uncertainty on the unfolded spectra was estimated by shifting the energy scale by the corresponding error and comparing it with the nominal distribution. As we can see from table \ref{tab:systematics}, the energy scale is the main source of uncertainty, especially in the high energy region where the slope of the distributions is steep. The systematic uncertainty strongly depends on energy, ranging from about 1\% to about 100\%.

\item Beam center
\enlargethispage{+1\baselineskip}

In order to measure the event pseudorapidity, we need to know the beam center, defined as the projection of the beam direction at the interaction point on the detector plane. This parameter was obtained by fitting the position distribution of high energy hadrons with a two dimensional function. The uncertainty estimated using this procedure was $\pm$0.3~mm, a value well consistent with the observed run-by-run fluctuations of this parameter. In order to evaluate the systematic uncertainty on the spectrum, we shifted the beam center position by this quantity on the x and y axes. Making use of these four additional distributions, we obtained the systematic uncertainty from the maximum deviation between each of them and the nominal one. The systematic uncertainty is generally below 5\%.

\item Position resolution

In order to simplify the unfolding procedure, deconvolution takes into account only the effects due to energy smearing, and not to position resolution. Although migration caused by position misreconstruction was corrected applying fake and missed events correction factors, we decided to consider this effect as a source of systematic uncertainty as well. This contribution was estimated from simulations, comparing the spectrum obtained using reconstructed position with the one obtained using true position. After having evaluated this ratio with both QGSJET II-04 and EPOS-LHC, the maximum deviation found was used as an estimation of the systematic uncertainty. The systematic uncertainty is generally below 10\%.

\item Integrated luminosity

The error on the integrated luminosity, derived from ATLAS measurements taking into account LHCf data acquisition live time, was estimated to be 1.9\%: this value was considered as an energy independent systematic uncertainty.

\item Correction factors

The systematic uncertainties due to correction factors were discussed in section \ref{sec:corrections}: they are significant for the PID, multihit, and hadron contamination corrections, but negligible for beam-related background, fake events, and missed events corrections.

\item Interaction cross section

The main contribution to missed events correction factors is due to detection inefficiency, which, in turn, depends on the cross section that describes the interaction of a hadron with the detector. This parameter has similar values in the DPMJET 3.04 and QGSJET II-04 implementations inside EPICS, but it changes if we consider a different generator implemented inside another simulation toolkit. In order to estimate the uncertainty on this parameter, we compared the proton-tungsten interaction cross section used for the DPMJET 3.04 model in EPICS with the one obtained from the QGSP\_BERT 4.0 model in GEANT4 \cite{ref:GEANT4}, a completely different simulation toolkit. The maximum deviation found below 6.5~TeV is 6\%, and is considered as an energy independent systematic uncertainty on the spectrum. 

\item Unfolding

Four different terms contribute to the systematic uncertainty related to the unfolding procedure itself.
The first, here called \textit{unfolding method uncertainty} (mostly below 20\%), depends on the ability of the iterative Bayesian method to produce a spectrum consistent with the true level one. We estimated this term using simulations relative to the QGSJET II-04 and EPOS-LHC generators, unfolding the distributions making use of a flat prior and comparing the unfolded spectra obtained in this way with the true distribution. 
The second, here called \textit{unfolding prior uncertainty} (mostly below 20\%), derives from the dependence of the unfolded spectrum on the chosen prior used as the initial hypothesis of the iterative procedure. We estimated this term by comparing the unfolded results obtained using a flat prior with the ones obtained using priors chosen from the QGSJET II-04, EPOS-LHC and DPMJET 3.04 generators. As discussed in appendix \ref{app:isr}, the best knowledge we could have on the prior shape would be given by the assumption of $\mathrm{x_{F}}$ scaling \cite{ref:xF} and the  extrapolation from ISR measurements \cite{ref:ISR1, ref:ISR2}, using the method suggested in reference \cite{ref:PHENIX}.\footnote{
The Feynman variable is defined as $\mathrm{x_{F} \equiv 2 p_{Z}/\sqrt{s}}$, where $\mathrm{p_{Z}}$, the longitudinal component of the momentum, can be approximated to the energy $\mathrm{E}$ in the very forward region. The $\mathrm{x_{F}}$ scaling hypothesis asserts that,  for $\mathrm{x_{F}}$ above some value (typically $\mathrm{x_{F}\gtrsim0.20}$), the production cross section of secondary particles expressed as a function of $\mathrm{x_{F}}$ should be independent of $\mathrm{\sqrt{s}}$.
}
However, in the end we chose not to follow this approach as it introduces an unacceptably large unfolding bias. The third, here called \textit{unfolding interaction uncertainty} (up to 70\%), depends on the choice of the hadronic interaction model used to construct the response matrix. We estimated this term by comparing the unfolded results obtained simulating the detector interaction using two different models: DPMJET 3.04, employed for all simulations described in section \ref{sec:mc}, and QGSJET II-04. We decided against superimposing this uncertainty on the final results because, as described in appendix \ref{app:sps}, the large effects found are mainly due to the poor agreement between QGSJET II-04 and the experimental data obtained in various beam tests, which are anyway well reproduced by DPMJET 3.04. The fourth, here called \textit{unfolding algorithm uncertainty} (up to 100\%), derives from the different unfolded spectrum that we can obtain using different algorithms. We estimated this term comparing the unfolded results obtained using the iterative Bayesian method with those obtained using SVD \cite{ref:SVD} and TUnfold \cite{ref:TUnfold} (also implemented inside the RooUnfold package). In principle, this uncertainty should already be included in the first two, but we decided to conservatively consider it as an additional independent source of error. This is because the algorithm performances, tested using simulations, could change unpredictably when applied to real data, given the marked differences with the experimental results shown by all models, especially in Region A.

\end{itemize}

\section{Results}
\label{sec:results}

In this section, we present the measurements relative to neutron production and we compare them with model predictions, both before and after unfolding. Distributions are expressed in terms of the inclusive differential production cross section $\mathrm{d\sigma_{n}/dE}$, which was corrected for the limited coverage of the azimuthal angle. In the comparison with generators, for each model we used its own inelastic cross section, as reported in reference \cite{ref:LHCf_13TeV}.  

\begin{figure*}[tbp]
 \centering
 \includegraphics[width=.333\textwidth]{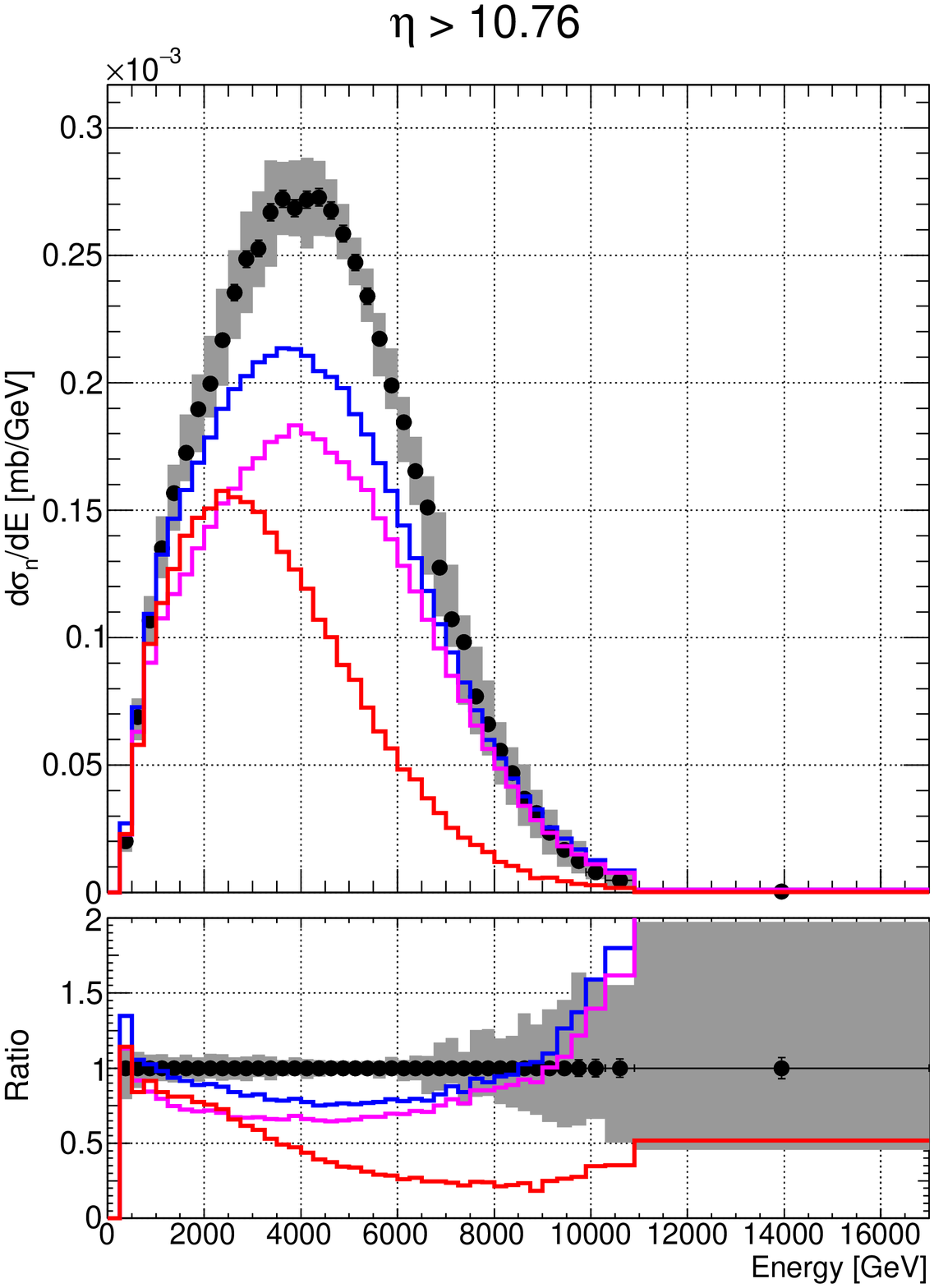}%
 \includegraphics[width=.333\textwidth]{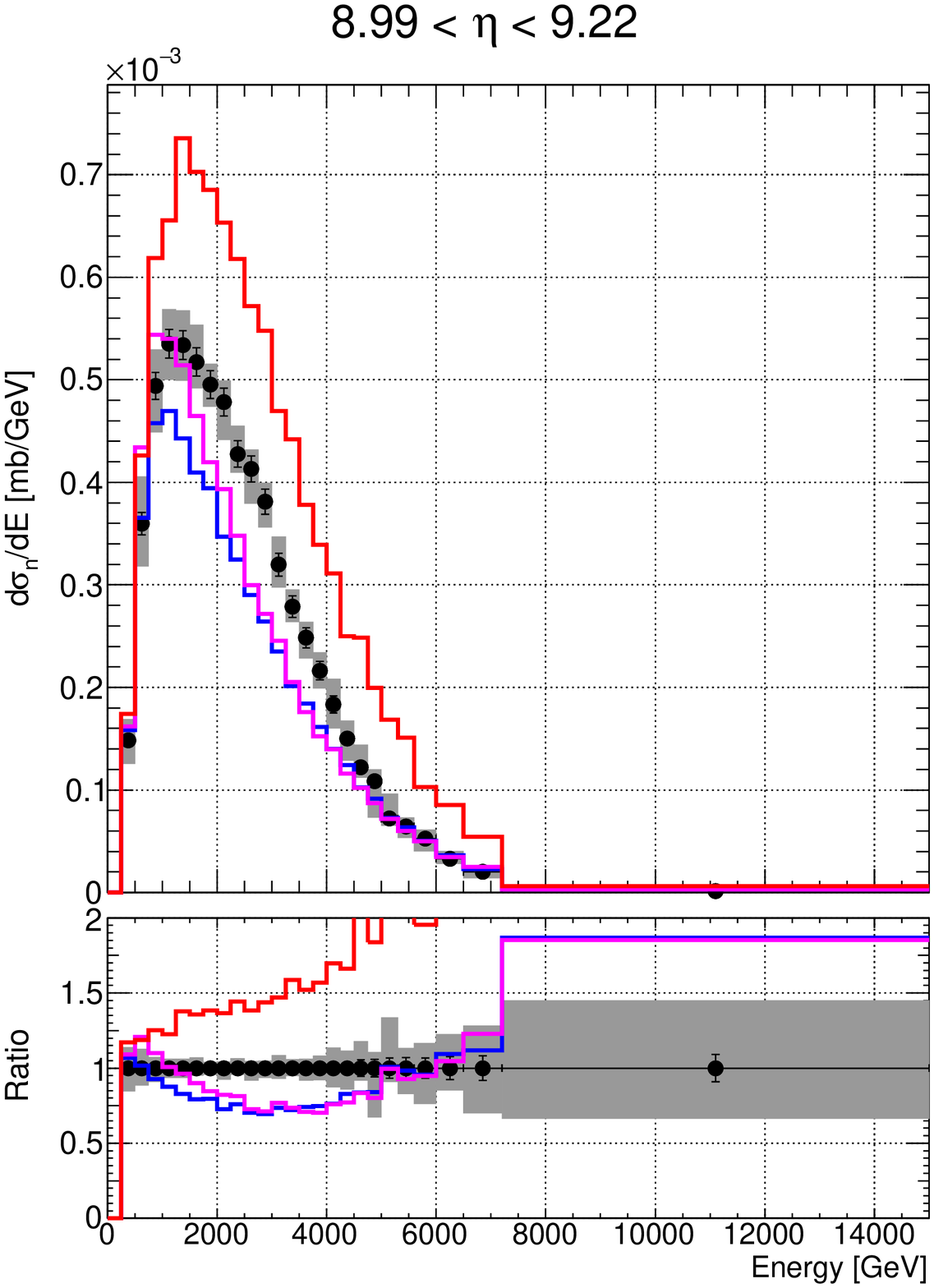}%
 \includegraphics[width=.333\textwidth]{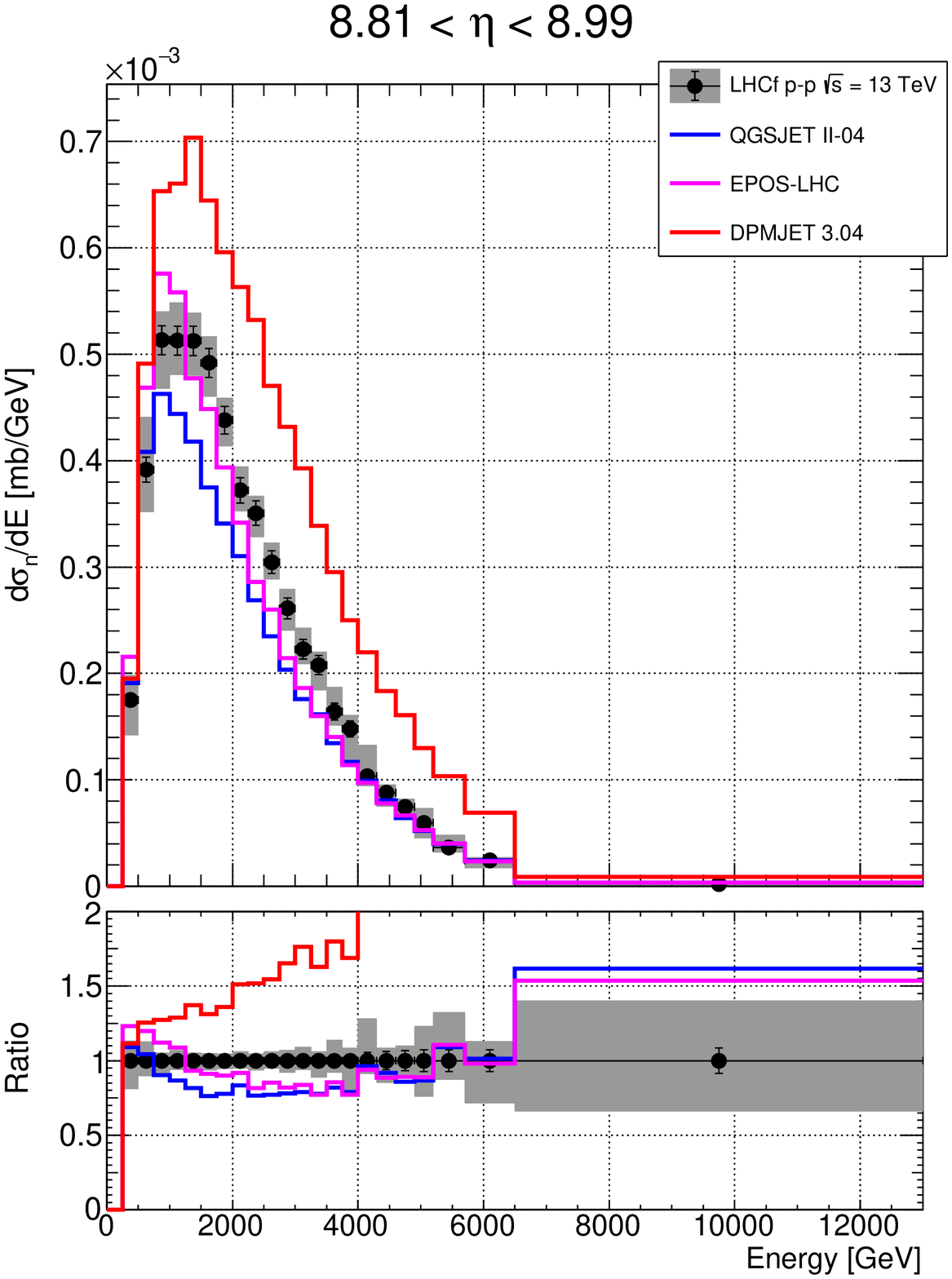}%
 \caption{Folded differential neutron production cross section for p-p collisions at $\sqrt{s} = 13$~TeV, measured using the LHCf Arm2 detector. Black markers represent the experimental data with statistical errors, whereas gray bands represent the quadratic sum of statistical and systematic uncertainties. Colored histograms refer to models predictions at the detector level. The top plot shows the energy distributions expressed as $\mathrm{d\sigma_{n}/dE}$ and the bottom one the ratios of these distributions to the experimental data points.}
\label{fig:folded}
\end{figure*}

The Arm2 $\mathrm{d\sigma_{n}/dE}$ folded distribution is shown in figure \ref{fig:folded}. This result is obtained before the application of the multihit correction factors, thus the total uncertainty is given by the quadratic sum of all statistical and systematic contributions relative to the folded spectrum, except multihit and interaction cross section errors. The binning was chosen so that the statistical uncertainty stays always below 10\%. As we can see, no model reproduces completely the experimental measurements and, in particular, the agreement of DPMJET 3.04 with data is very limited both in terms of spectral shape and total yield. For this reason, as described in section \ref{sec:mc}, we excluded this generator from the estimation of model dependent corrections and uncertainties.  

\begin{figure*}[tbp]
 \centering
 \includegraphics[width=.333\textwidth]{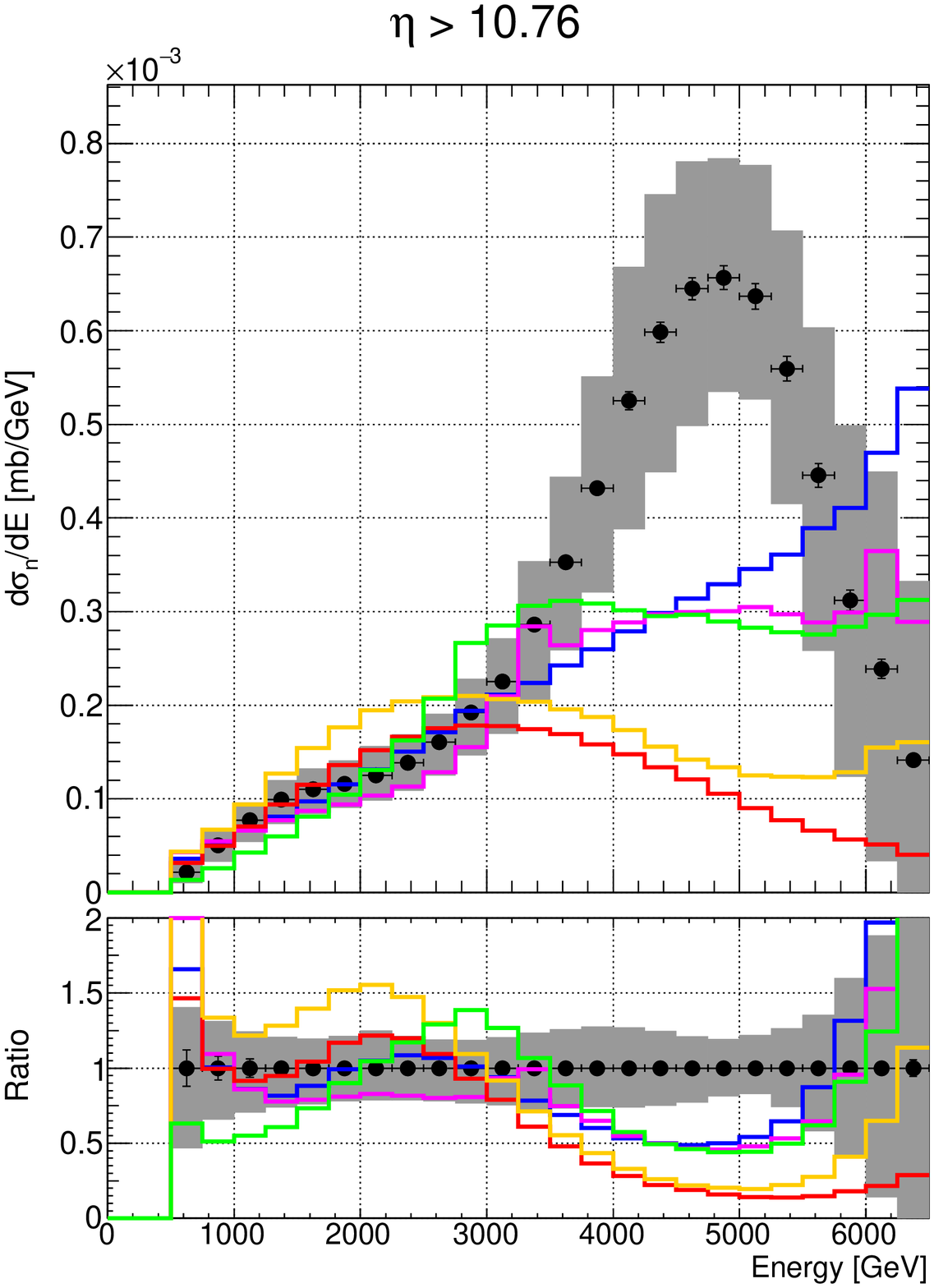}%
 \includegraphics[width=.333\textwidth]{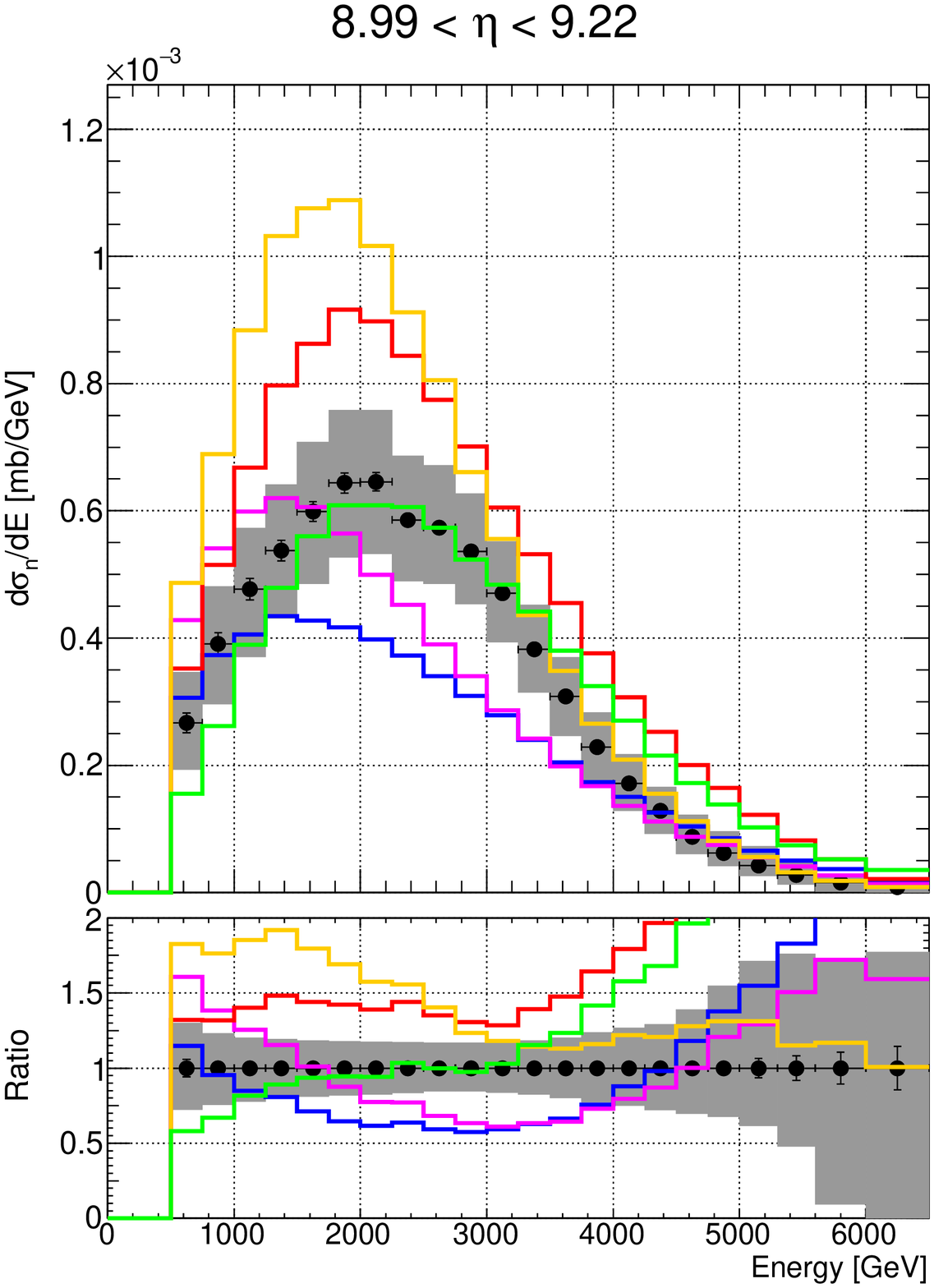}%
 \includegraphics[width=.333\textwidth]{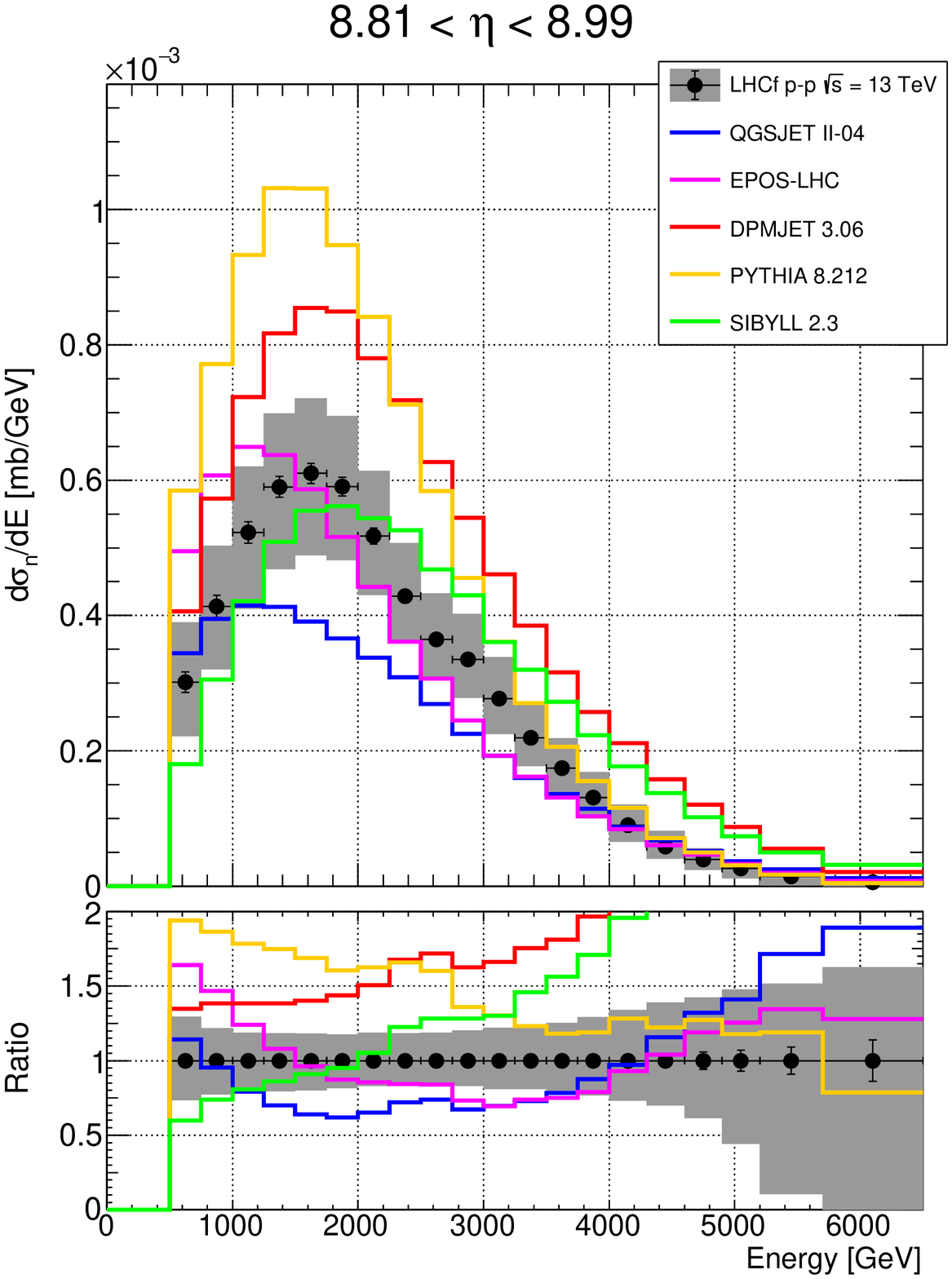}%
 \caption{Unfolded differential neutron production cross section for p-p collisions at $\sqrt{s} = 13$~TeV, measured using the LHCf Arm2 detector. Black markers represent the experimental data with statistical errors, whereas gray bands represent the quadratic sum of statistical and systematic uncertainties. Colored histograms refer to models predictions at the generator level. The top plot shows the energy distributions expressed as $\mathrm{d\sigma_{n}/dE}$ and the bottom one the ratios of these distributions to the experimental data points.}
\label{fig:unfolded}
\end{figure*}

\enlargethispage{+2\baselineskip}
The Arm2 $\mathrm{d\sigma_{n}/dE}$ unfolded distribution is shown in figure \ref{fig:unfolded}. The total uncertainty is given by the quadratic sum of all statistical and systematic contributions. The values of the inclusive differential production cross section are also summarized in appendix \ref{app:cross_section}. In the pseudorapidity region $\eta > 10.76$, neutron production exhibits a peak structure at around 5~TeV, which is not predicted by any of the models available. As discussed in appendix \ref{app:isr}, the peak is present independently of the choice of a flat prior or even an ISR prior, though the position of the peak itself does change from $\mathrm{x_{F}} \sim 0.75$ to $\mathrm{x_{F}} \sim 0.80$, which, in turn, does affect the conclusions on the validity of the $\mathrm{x_{F}}$ scaling hypothesis. Regarding the models, QGSJET II-04 shows a constant increase in the production rate up to high energies, whereas EPOS-LHC and SIBYLL 2.3 reach an approximately constant production rate above 3.5~TeV. These three generators do not contemplate the presence of a high energy peak in the forward neutron differential cross section, leading to a lower yield at 5~TeV and a higher yield at 6.5~TeV. On the other hand, DPMJET 3.06 and PYTHIA 8.212 show a peak structure, but they strongly underestimate both the position of the peak and the production rate at high energies. For all generators the total production cross section at $\eta > 10.76$ is lower than the one experimentally observed: QGSJET II-04 is the model having the smallest deficit with respect to data, amounting to a value of around 20\%. Even in the pseudorapidity regions $8.99 < \eta < 9.22$ and $8.81 < \eta < 8.99$, no model reproduces completely the experimental measurements, although deviations are smaller than at $\eta > 10.76$. For these two pseudorapidity intervals, SIBYLL 2.3 and EPOS-LHC show the best overall agreement with our data at $8.99 < \eta < 9.22$ and $8.81 < \eta < 8.99$, respectively. In particular, SIBYLL 2.3 is compatible with the experimental measurements in the region between 1.5 TeV and 2.5~TeV, where neutron production is maximum, but it is softer below and harder above this interval. The other models generally underestimate (QGSJET II-04, EPOS-LHC) or overestimate (DPMJET 3.06, PYTHIA 8.212) the inclusive differential cross section in all the energy range.

\enlargethispage{+2\baselineskip}
The general trend of the experimental results is similar to what observed at $\mathrm{\sqrt{s}} = $ 7~TeV \cite{ref:LHCf_7TeV}. Direct comparison of the data-models agreement cannot be made because the versions used here are different with respect to the ones employed there. In particular, QGSJET II-04, EPOS-LHC and SIBYLL 2.3 were tuned using the LHC Run I results. Comparing the \textit{pre-LHC} and \textit{post-LHC} version of SIBYLL, we can observe a significant increase of the neutron production in all the pseudorapidity regions and this fact improves the agreement of the model with the current experimental measurements. Differently, QGSJET and EPOS are not affected by relevant changes. Whereas no significant variation is found also in PYTHIA, DPMJET exhibits a very different inclusive differential cross section in the two cases. Because no relevant change in the mechanisms responsible for forward neutron production is expected between $\mathrm{\sqrt{s}} =$ 7 and 13~TeV, this variation must be due to the fact that the $\mathrm{p_{T}}$ coverage is different in the two LHC runs, i.e. the $\mathrm{p_{T}}$ intervals corresponding to a given pseudorapidity region at 13 TeV are almost twice the ones at 7 TeV.

\section{Summary}

The LHCf experiment measured the inclusive differential cross section of forward neutrons produced in proton-proton collisions at $\sqrt{s} = 13$~TeV. The analysis covers the same three pseudorapidity regions considered in the previous results at $\sqrt{s} = 7$~TeV: $\eta > 10.76$, $8.99 < \eta < 9.22$ and $8.81 < \eta < 8.99$. Experimental measurements were compared to the prediction of several hadronic interaction models: QGSJET II-04, EPOS-LHC, SIBYLL 2.3, DPMJET 3.06 and PYTHIA 8.212. No one of these generators showed complete or acceptable agreement with our data, indicating that further progress must be made in the understanding of particle production in the forward region. For $\eta > 10.76$, the experimental energy distribution exhibits a peak structure at around 5~TeV that is not predicted by any of the models, which also  underestimate the total production cross section: QGSJET II-04 shows the smallest deficit with respect to data for the whole energy range. In the other two regions, the experimental energy distributions exhibit a peak structure at around 1.5-2.5~TeV that is well reproduced by SIBYLL 2.3: SIBYLL 2.3 and EPOS-LHC show the best overall agreement for $8.99 < \eta < 9.22$ and $8.81 < \eta < 8.99$, respectively.

The results presented in this paper are relative to the Arm2 detector only. In the future we plan to extend this analysis to the Arm1 detector as well, in order to enlarge the pseudorapidity coverage, exploiting the slightly different acceptance of the two detectors. An additional improvement of our results will be possible thanks to the common data acquisition that the LHCf and the ATLAS experiments had during the operations relative to the data analyzed in this paper. By exploiting the ATLAS information in the central region, we can tag LHCf triggers on an event-by-event basis, investigating the different mechanisms responsible for neutron production in the forward region \cite{ref:Diffractive_Zhou, ref:ATLAS_LHCf_note}.

\acknowledgments
We thank the CERN staff and ATLAS Collaboration for their essential contributions to the successful operation of LHCf. We are grateful to S. Ostapchenko for useful comments about QGSJET II-04 generator and to C. Baus, T. Pierog, and R. Ulrich for the implementation of the CRMC interface tool. This work was supported by the Japanese Society for the Promotion of Science (JSPS) KAKENHI (Grant Numbers JP26247037, JP23340076) in Japan, by Istituto Nazionale di Fisica Nucleare (INFN) in Italy and by the joint research program of the Institute for Cosmic Ray Research (ICRR), University of Tokyo. Parts of this work were performed using the computer resource provided by ICRR (University of Tokyo), CERN and CNAF (INFN).

\clearpage

\appendix
\section{Dependence of the unfolded result on the choice of flat or ISR prior}
\label{app:isr}

\begin{figure}[tbp]
\centering
\includegraphics[width=0.475\linewidth]{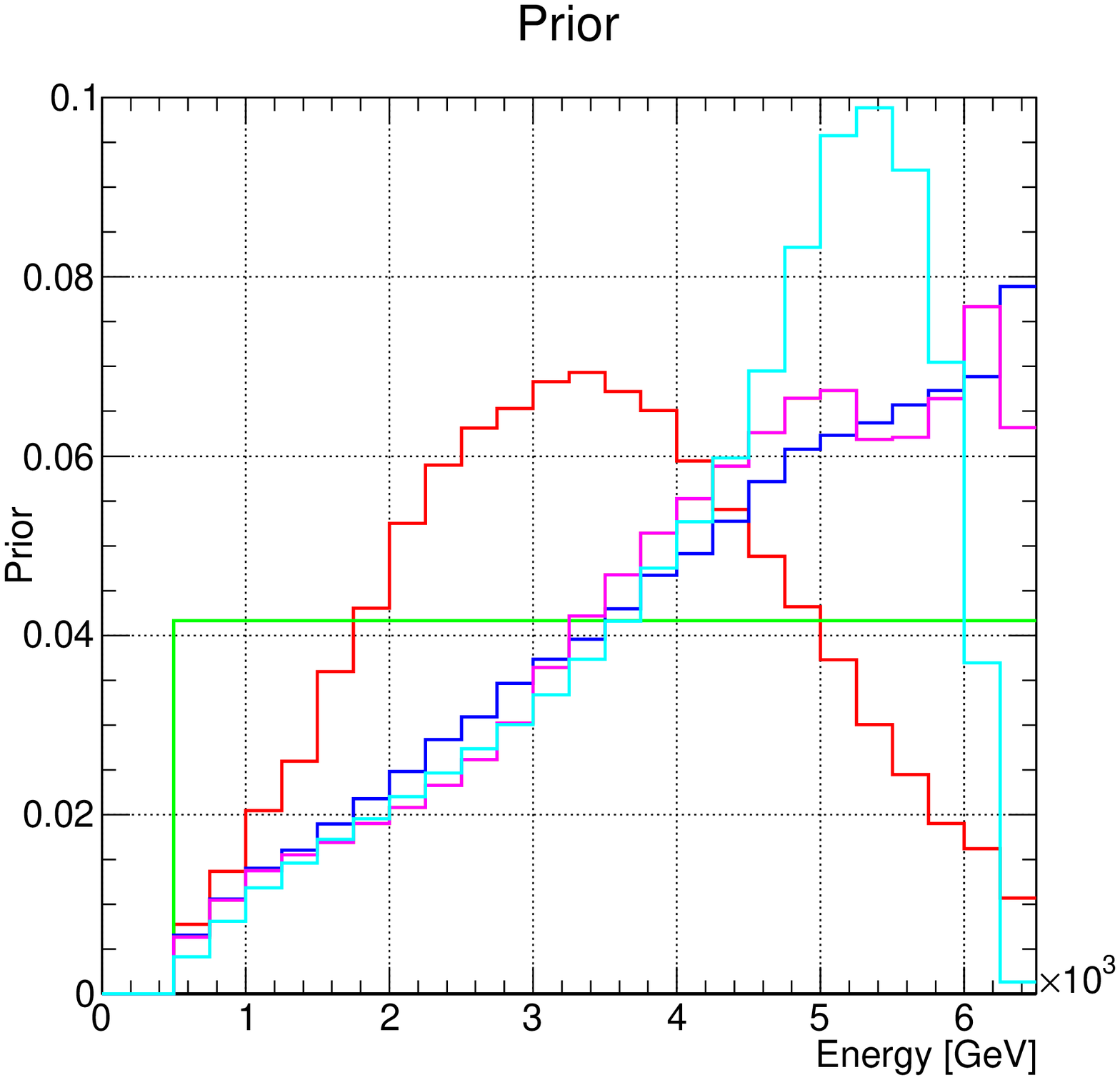}
\hspace{10mm}%
\includegraphics[width=0.475\linewidth]{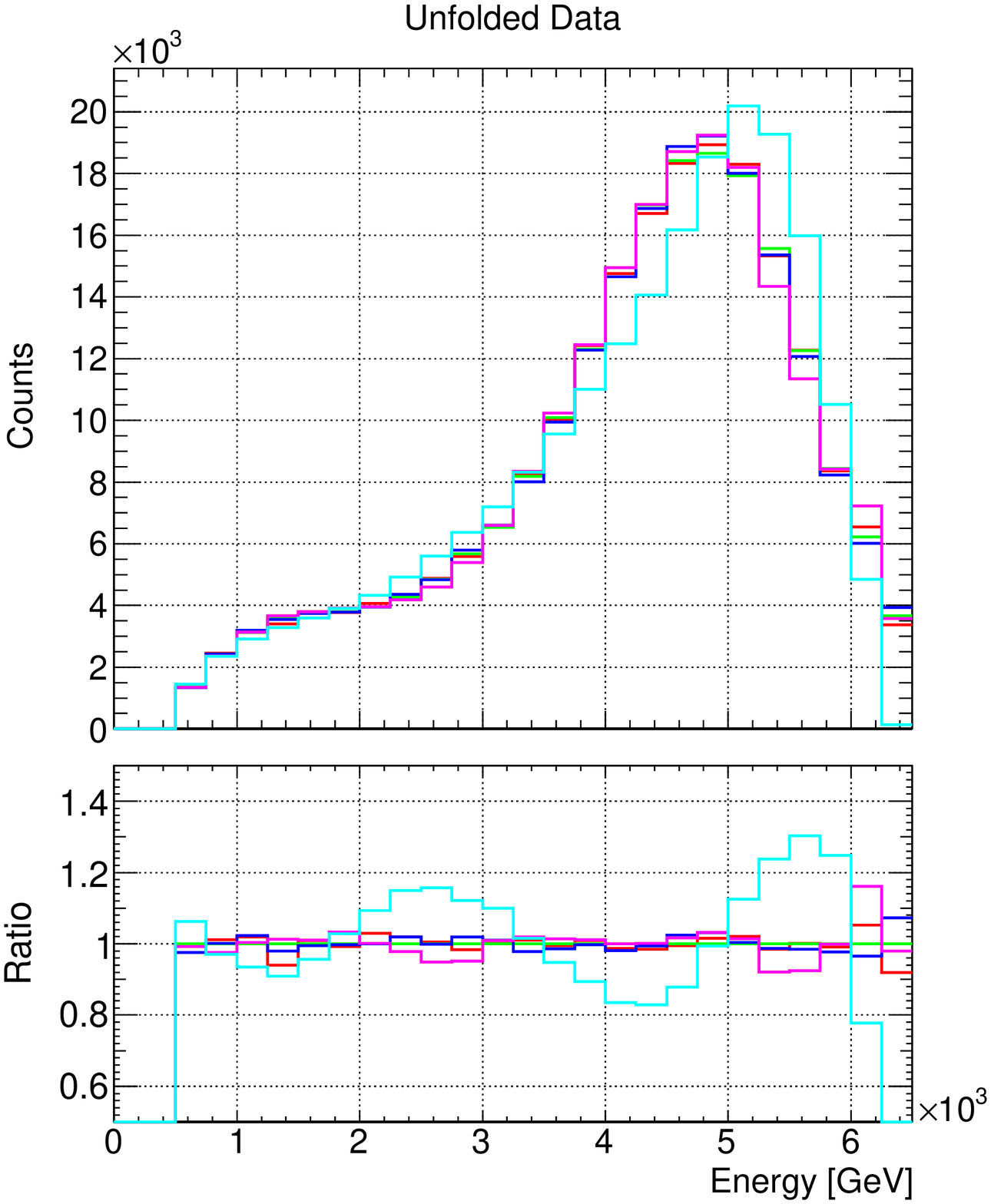}
\includegraphics[width=0.475\linewidth]{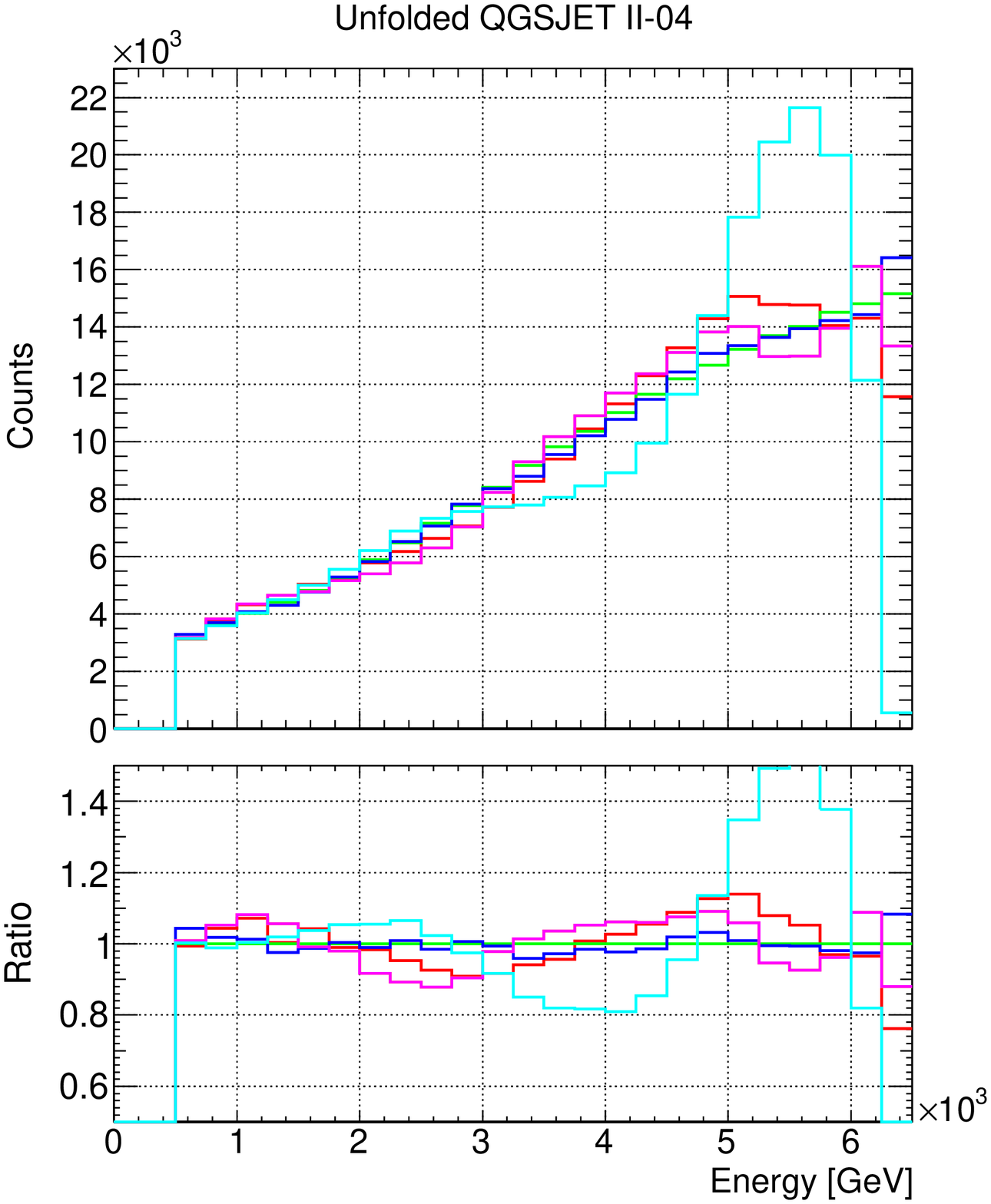}
\caption{Top figure shows the different priors used for iterative Bayesian unfolding and bottom figures show the unfolded results corresponding to these different priors: left is  experimental measurements, right is QGSJET II-04 simulations. The unfolded results are shown both as absolute value (top) and as the ratio to the spectrum ontained using flat prior (bottom). The different priors are: flat (green), QGSJET II-04 (blue), EPOS-LHC (magenta), DPMJET 3.04 (red), ISR (cyan). The ISR prior was extrapolated applying the method described in reference \cite{ref:PHENIX} to the data relative to p-p collisions at $\sqrt{s} = 44.9$~GeV \cite{ref:ISR1, ref:ISR2}. Note that the ISR prior leads to an unfolded result that is very different from the one obtained with the other priors.}
\label{fig:isr_bias}
\end{figure}

Experimental measurements of inclusive differential neutron production cross section in the forward region have been carried out at ISR \cite{ref:ISR1, ref:ISR2}, and at RHIC by the PHENIX experiment \cite{ref:PHENIX}. The different results are in good agreement between each other and demonstrates the validity of $\mathrm{x_{F}}$ scaling hypothesis for p-p collisions in the range $\sqrt{s} = 30.6-200$~GeV. Thus, when choosing the prior used as initial hypothesis of the iterative Bayesian unfolding, it would be in principle better to choose the one extrapolated from ISR measurements rather than the flat one. This extrapolation can be easily done using the same method described in reference \cite{ref:PHENIX}, limiting the integration to the acceptance of Region A in LHCf, corresponding to $\mathrm{p_{T} < 0.276 ~ GeV/c \cdot x_{F} }$ in the case of p-p collisions at $\sqrt{s} = 13$~TeV. Here we call \textit{ISR prior} the one obtained extrapolating the ISR results relative to p-p collisions at $\sqrt{s} = 44.9$~GeV according to the procedure mentioned above. Unfortunately, as shown in figure \ref{fig:isr_bias}, this prior was found to have a large bias on the final result, observed both on experimental data and generator simulations. In particular, being very different from all models, the ISR prior leads to an unfolded spectrum that is very different from the generator truth in the case of QGSJET II-04. This is likely due to the spectral shape of the prior, which shows a sharp peak at about $\mathrm{x_{F}} = ~ 0.80$ that rapidly decreases to 0 at $\mathrm{x_{F}} = ~ 1$. In addition, this problem was found only with the ISR prior, but not with an artificial prior obtained shifting the peak position from $\mathrm{x_{F}} = ~ 0.80$ towards smaller or larger values. Thus, we concluded that the ISR prior is not a safe choice for our analysis because, in the case the distribution present in nature was different from ISR measurements, the unfolding procedure would lead to a result that is different from the true one. As shown in figure \ref{fig:isr_bias}, all other priors lead to similar results and, in any case, the residual difference is considered as a source of systematic uncertainty on our final measurements. Note that all this discussion applies only to Region A, but not to Region B and C, whose $\mathrm{p_{T}}$ coverage is outside the validity of the extrapolation suggested in reference \cite{ref:PHENIX}.

\clearpage

\section{Test of hadronic interaction models using beam test data}
\label{app:sps}

\begin{figure*}[tbp]
\centering
\includegraphics[width=0.5\linewidth]{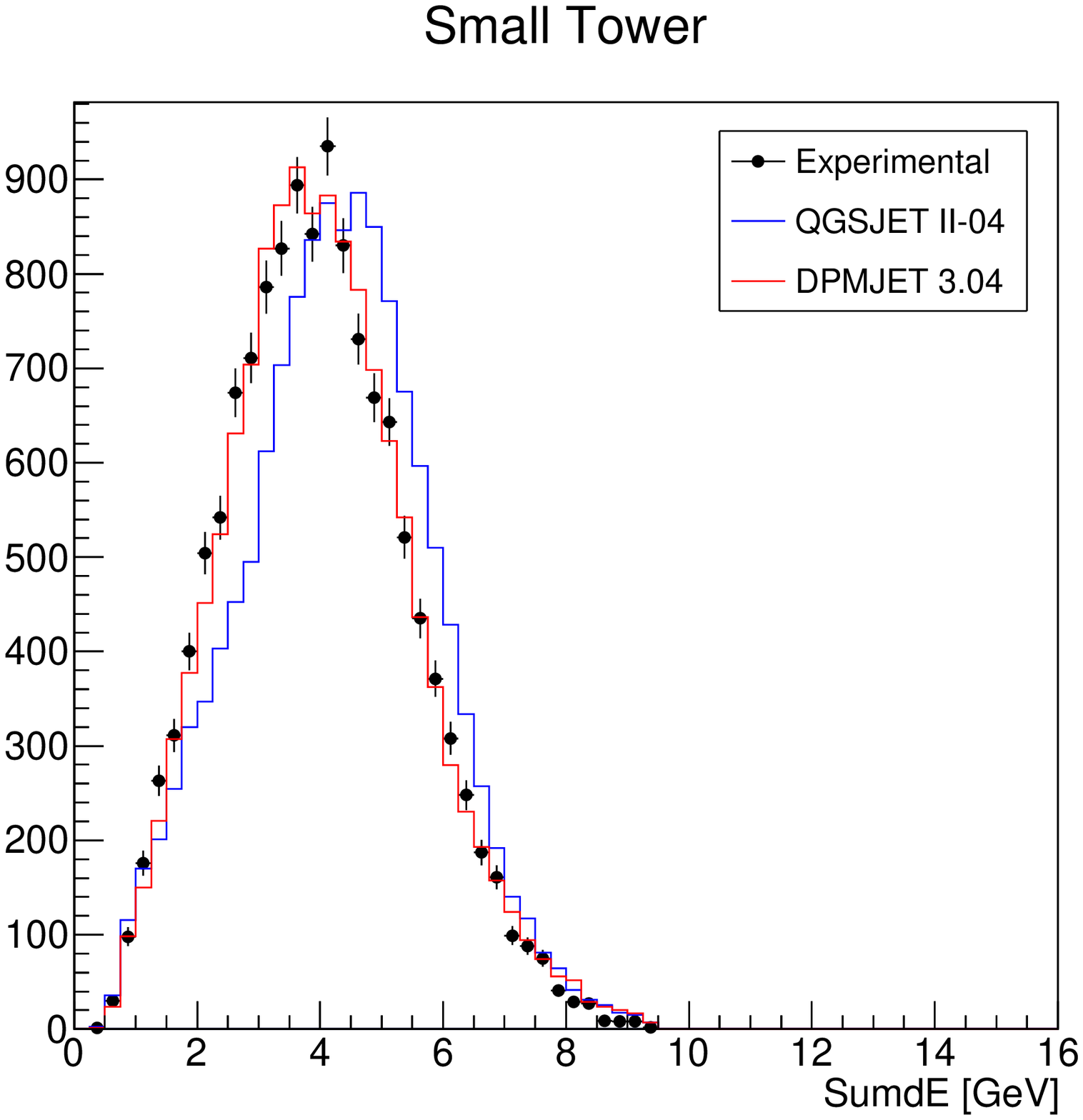}%
\includegraphics[width=0.5\linewidth]{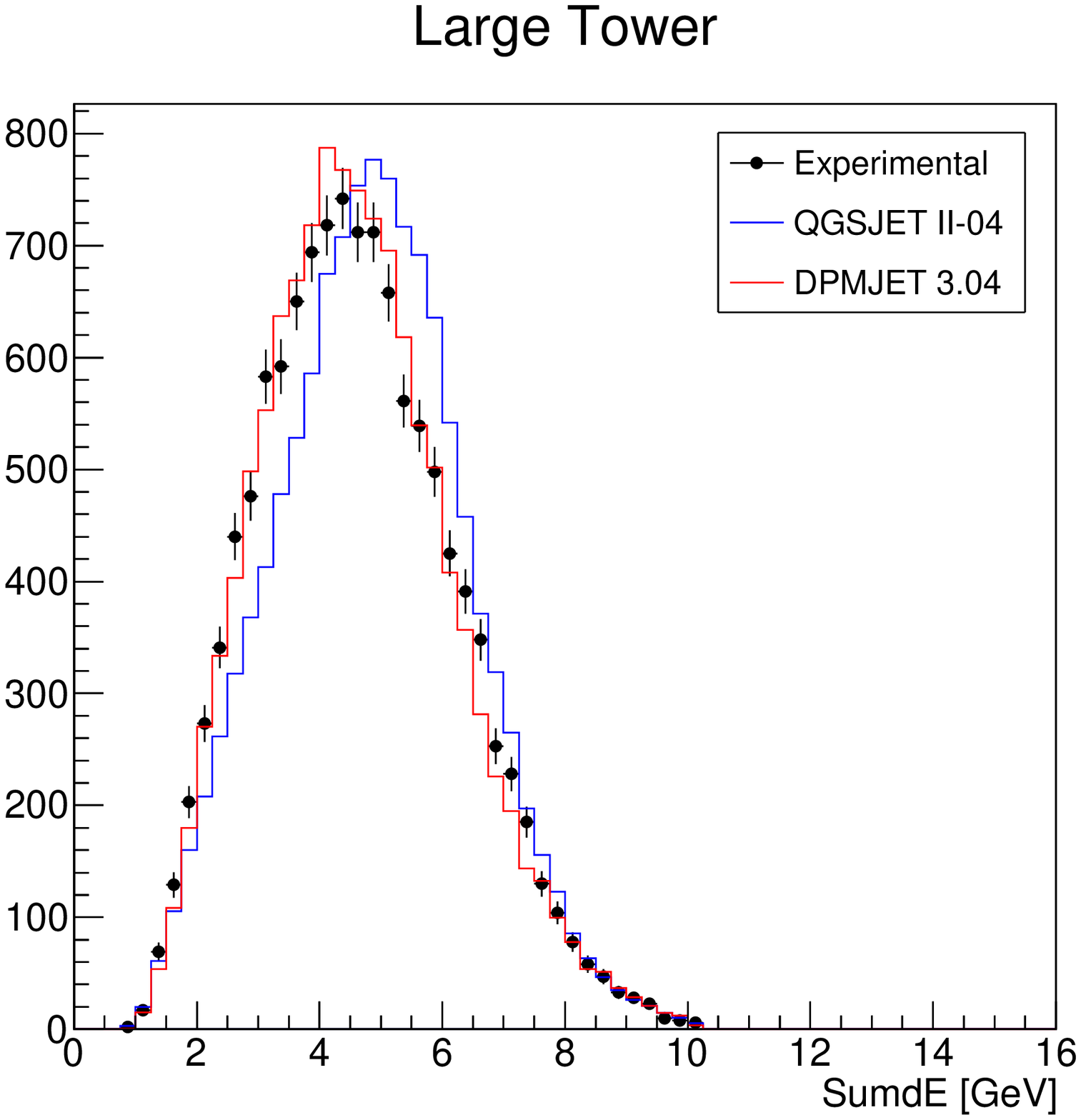}%
\caption{Total energy deposit distributions relative to 350~GeV protons. Black points with error bars are experimental data with statistical uncertainty. Histograms refer to DPMJET 3.04 and QGSJET II-04 predictions at the detector level. Left is small tower, right is large tower.}
\label{fig:sps}
\end{figure*}

The choice of the hadronic interaction model used to simulate the detector response plays a very important role in the analysis described in this paper. It is employed for detector calibration, for the estimation of correction factors, for the evaluation of systematic uncertainties, and for spectra unfolding. Unfortunately, the agreement between different models is not good in the energy range considered in this analysis. Because of this reason, we decided to test them using experimental data at the highest energy achievable at beam test facilities. These data were acquired at the CERN Super Proton Synchrotron (SPS) in July-August 2015, making use of muon, electron and proton beams at different energies. The total energy deposit of 350 GeV protons in the detector was compared to the one predicted by DPMJET 3.04 and QGSJET II-04 models in the same experimental configuration. The event selection applied to these distributions differs from the one described in section \ref{sec:reconstruction} for two main reasons: the reduction of the software trigger threshold (75~MeV for small tower, 150~MeV for large tower) to cope with the otherwise small detection efficiency at this energy; the reduction of the detector fiducial area to a small square around center ($\mathrm{5~mm \times 5~mm}$ for small tower, $\mathrm{6~mm \times 6~mm}$ for large tower) in order to select a region of uniform response. The comparison between data and MC is shown in figure \ref{fig:sps}, whereas the parameters of the distributions are reported in table \ref{tab:sps}. As we can see, DPMJET 3.04 is in very good agreement with experimental measurements, whereas QGSJET II-04 leads to an energy deposit that is slightly higher than the one observed. Thus, in the current analysis, we decided to use the DPMJET 3.04 model every time we need to simulate the interaction of a particle with the detector. In particular, the deviation of the energy mean and the energy resolution between DPMJET 3.04 predictions and experimental measurements is at most 2\%. Because this difference is compatible with the error on the gain of scintillator channels, we decided to neglect this additional contribution to the total systematic uncertainty on the energy scale. As a final remark, it is important to note that the choice of the model used to simulate the interaction with the detector is independent from the ability of the same model to reproduce the distributions of secondary particles produced in $\sqrt{s} = 13$~TeV p-p collisions. This is because, apart from the fact that in the calorimeter the first interaction is between the incoming particle and tungsten, there is a difference of more than four orders of magnitude between the highest possible energy of a particle reaching the detector (6.5~TeV) and the energy of the primary proton in the laboratory frame where the other proton is at rest (90~PeV).

\begin{table}[tbp]
  \begin{center}
    \begin{tabular}{|ll|c|c|c|c|}
  	\hline
	 & & Mean [GeV] & Ratio & $\sigma$/Mean [$\%$] & Ratio \\
	\hline
	 & Experiment	&3.965 &- &37.824 &-\\ 
	 Small tower & QGSJET II-04 &4.276 &1.078 &36.166	&0.956\\
	 & DPMJET 3.04    &4.023 &1.015 &37.754	&0.998\\
	\hline
	 & Experiment	&4.620 &- &34.503 &-\\ 
	 Large Tower & QGSJET II-04 &4.864 &1.053 &32.077 &0.927\\
	 & DPMJET 3.04     &4.601 &0.996 &33.746 &0.978\\
 	\hline
  \end{tabular}
  \caption{Energy mean and energy resolution for distributions shown in figure \ref{fig:sps}. For each parameter, the ratio of model prediction to the experimental value is also reported.}
  \label{tab:sps}
  \end{center}
\end{table}

\clearpage

\section{Table of inclusive differential production cross section}
\label{app:cross_section}

\begin{table} [htbp]
  \scriptsize
  \begin{center}
    \begin{tabular}{|P{1.5cm}|P{2.65cm}|P{1.5cm}|P{2.65cm}|P{1.5cm}|P{2.65cm}|}
      \hline
      \multicolumn{2}{|c}{} & \multicolumn{2}{|c|}{} & \multicolumn{2}{c|}{} \\
      \multicolumn{2}{|c}{\boldmath$\eta > 10.76$} & \multicolumn{2}{|c|}{\boldmath$8.99 < \eta < 9.22$} & \multicolumn{2}{c|}{\boldmath$8.81 < \eta < 8.99$} \\
      \multicolumn{2}{|c}{} & \multicolumn{2}{|c|}{} & \multicolumn{2}{c|}{} \\
      \hline
      &  &  &  &  &\\
      \textbf{Energy [GeV]} & \boldmath$\mathrm{d\sigma_{n}/dE}$ \textbf{[mb/GeV]} & \textbf{Energy [GeV]} & \boldmath$\mathrm{d\sigma_{n}/dE}$ \textbf{[mb/GeV]}
      & \textbf{Energy [GeV]} & \boldmath$\mathrm{d\sigma_{n}/dE}$ \textbf{[mb/GeV]} \\
      &  &  &  &  &\\
      \hline
      &  &  &  &  &\\
      500--750 & $(2.16_{-1.15}^{+0.88}) \times 10^{-5}$  &       500--750 & $(2.67_{-0.74}^{+0.80}) \times 10^{-4}$  &       500--750 & $(3.02_{-0.80}^{+0.88}) \times 10^{-4}$ \\
      750--1000 & $(5.01_{-1.72}^{+1.57}) \times 10^{-5}$  &       750--1000 & $(3.91_{-0.95}^{+0.90}) \times 10^{-4}$  &       750--1000 & $(4.14_{-0.93}^{+0.89}) \times 10^{-4}$ \\
      1000--1250 & $(7.70_{-2.27}^{+1.87}) \times 10^{-5}$  &       1000--1250 & $(4.77_{-1.07}^{+0.96}) \times 10^{-4}$  &       1000--1250 & $(5.23_{-1.12}^{+0.97}) \times 10^{-4}$ \\
      1250--1500 & $(9.92_{-2.60}^{+2.05}) \times 10^{-5}$  &       1250--1500 & $(5.38_{-1.06}^{+1.04}) \times 10^{-4}$  &       1250--1500 & $(5.90_{-1.22}^{+1.08}) \times 10^{-4}$ \\
      1500--1750 & $(1.10_{-0.27}^{+0.22}) \times 10^{-4}$  &       1500--1750 & $(5.99_{-1.13}^{+1.09}) \times 10^{-4}$  &       1500--1750 & $(6.10_{-1.21}^{+1.11}) \times 10^{-4}$ \\
      1750--2000 & $(1.16_{-0.25}^{+0.25}) \times 10^{-4}$  &       1750--2000 & $(6.44_{-1.17}^{+1.15}) \times 10^{-4}$  &       1750--2000 & $(5.91_{-1.09}^{+1.04}) \times 10^{-4}$ \\
      2000--2250 & $(1.25_{-0.27}^{+0.31}) \times 10^{-4}$  &       2000--2250 & $(6.46_{-1.13}^{+1.13}) \times 10^{-4}$  &       2000--2250 & $(5.18_{-0.87}^{+0.96}) \times 10^{-4}$ \\
      2250--2500 & $(1.39_{-0.30}^{+0.30}) \times 10^{-4}$  &       2250--2500 & $(5.86_{-0.96}^{+1.01}) \times 10^{-4}$  &       2250--2500 & $(4.29_{-0.66}^{+0.79}) \times 10^{-4}$ \\
      2500--2750 & $(1.60_{-0.35}^{+0.30}) \times 10^{-4}$  &       2500--2750 & $(5.74_{-0.89}^{+0.98}) \times 10^{-4}$  &       2500--2750 & $(3.65_{-0.59}^{+0.68}) \times 10^{-4}$ \\
      2750--3000 & $(1.92_{-0.45}^{+0.36}) \times 10^{-4}$  &       2750--3000 & $(5.36_{-0.83}^{+0.91}) \times 10^{-4}$  &       2750--3000 & $(3.35_{-0.56}^{+0.67}) \times 10^{-4}$ \\
      3000--3250 & $(2.25_{-0.56}^{+0.46}) \times 10^{-4}$  &       3000--3250 & $(4.71_{-0.77}^{+0.81}) \times 10^{-4}$  &       3000--3250 & $(2.78_{-0.52}^{+0.61}) \times 10^{-4}$ \\
      3250--3500 & $(2.86_{-0.83}^{+0.67}) \times 10^{-4}$  &       3250--3500 & $(3.82_{-0.67}^{+0.70}) \times 10^{-4}$  &       3250--3500 & $(2.19_{-0.42}^{+0.51}) \times 10^{-4}$ \\
      3500--3750 & $(3.53_{-0.94}^{+0.91}) \times 10^{-4}$  &       3500--3750 & $(3.08_{-0.62}^{+0.62}) \times 10^{-4}$  &       3500--3750 & $(1.74_{-0.35}^{+0.44}) \times 10^{-4}$ \\
      3750--4000 & $(4.32_{-1.11}^{+1.19}) \times 10^{-4}$  &       3750--4000 & $(2.29_{-0.53}^{+0.54}) \times 10^{-4}$  &       3750--4000 & $(1.31_{-0.30}^{+0.38}) \times 10^{-4}$ \\
      4000--4250 & $(5.25_{-1.37}^{+1.43}) \times 10^{-4}$  &       4000--4250 & $(1.71_{-0.43}^{+0.46}) \times 10^{-4}$  &       4000--4300 & $(9.03_{-2.42}^{+3.05}) \times 10^{-5}$ \\
      4250--4500 & $(5.98_{-1.49}^{+1.47}) \times 10^{-4}$  &       4250--4500 & $(1.28_{-0.36}^{+0.38}) \times 10^{-4}$  &       4300--4600 & $(5.85_{-1.76}^{+2.28}) \times 10^{-5}$ \\
      4500--4750 & $(6.45_{-1.47}^{+1.36}) \times 10^{-4}$  &       4500--4750 & $(8.77_{-2.66}^{+3.41}) \times 10^{-5}$  &       4600--4900 & $(3.95_{-1.52}^{+1.67}) \times 10^{-5}$ \\
      4750--5000 & $(6.57_{-1.22}^{+1.27}) \times 10^{-4}$  &       4750--5000 & $(6.21_{-2.02}^{+3.41}) \times 10^{-5}$  &       4900--5200 & $(2.64_{-1.47}^{+1.25}) \times 10^{-5}$ \\
      5000--5250 & $(6.37_{-1.10}^{+1.40}) \times 10^{-4}$  &       5000--5300 & $(4.24_{-1.63}^{+3.02}) \times 10^{-5}$  &       5200--5700 & $(1.46_{-1.30}^{+0.75}) \times 10^{-5}$ \\
      5250--5500 & $(5.59_{-1.45}^{+1.47}) \times 10^{-4}$  &       5300--5600 & $(2.74_{-1.43}^{+2.08}) \times 10^{-5}$  &       5700--6500 & $(6.23_{-6.23}^{+3.90}) \times 10^{-6}$ \\
      5500--5750 & $(4.46_{-1.87}^{+1.58}) \times 10^{-4}$  &       5600--6000 & $(1.56_{-1.42}^{+1.12}) \times 10^{-5}$  &  & \\
      5750--6000 & $(3.12_{-1.88}^{+1.87}) \times 10^{-4}$  &       6000--6500 & $(8.39_{-8.39}^{+6.46}) \times 10^{-6}$  &  & \\
      6000--6250 & $(2.39_{-2.05}^{+2.11}) \times 10^{-4}$  &  &  &  & \\
      6250--6500 & $(1.41_{-1.41}^{+1.91}) \times 10^{-4}$  &  &  &  & \\
      &  &  &  &  &\\
      \hline
    \end{tabular}
    \caption{Inclusive differential neutron production cross section for p-p collisions at $\sqrt{s} = 13$~TeV. The measurements refer to the results obtained using the LHCf Arm2 detector in the three pseudorapidity regions considered in the current analysis. Upper and lower uncertainties, expressed as the quadratic sum of statistical and systematics contributions, are also reported.}
    \label{tab:results}
  \end{center}
\end{table}

\end{document}